%% file: paperCP.tex
\documentstyle[12pt,epsfig,fleqn]{article}
 
\newcommand\sfrac[2]{{\textstyle \frac{#1}{#2}}}
\newcommand{\La}{{\mathcal L}} 
\newcommand{\mIm}{\,\mbox{\small $\Im$m\,}} 
\newcommand{\eRe}{\,\mbox{\small $\Re$e\,}} 
\newcommand{\be}{\begin{eqnarray}} 
\newcommand{\ee}{\end{eqnarray}} 
\newcommand{\nn}{\nonumber} 
\newcommand{\dgs}{d^{\gamma}{\!\scriptstyle (}s{\scriptstyle )}\,} 
\newcommand{\dzs}{d^{Z}{\!\scriptstyle (}s{\scriptstyle )}\,} 
\newcommand{\plr}{\stackrel{\leftrightarrow}{\partial}{}\!\!} 
\newcommand{\gsim}{\;\raisebox{-0.9ex} 
                     {$\textstyle\stackrel{\textstyle>}{\sim}$} \;}

\begin{document} 

\begin{titlepage}

\vspace*{-2cm}
\begin{flushright}
HEPHY-PUB 669/97\\
UWThPh-1997-13\\
hep-ph / 9705245
\end{flushright}

\begin{center}

\vspace{1cm}

\begin{Large} {\bf
  Electroweak dipole moment form factors \\[5mm]
  of the top quark in supersymmetry
}\end{Large}

\vspace{2cm}

{\large A. Bartl} \\
{\em Institut f\"ur Theoretische Physik, Universit\"at Wien, \\
     A-1090 Vienna, Austria} \\

\vspace{1cm}

{\large E. Christova} \\
{\em Institute of Nuclear Research and Nuclear Energy, \\
     Boul. Tzarigradsko Chaussee 72, Sofia 1784, Bulgaria} \\

\vspace{1cm}

{\large T. Gajdosik, W. Majerotto} \\
{\em Institut f\"ur Hochenergiephysik der 
     \"Osterreichischen Akademie der Wissenschaften, \\
     A-1050 Vienna, Austria}
\end{center}

\vfill

\begin{abstract}

The CP violating electric and weak dipole moment form 
factors of the top quark, $\dgs$ and $\dzs$, appear in the 
process \mbox{$e^{+}e^{-} \to t\,\bar{t}$}. We present a 
complete analysis of these dipole moment form factors within the 
Minimal Supersymmetric Standard Model with complex parameters. We 
include gluino, chargino, and neutralino exchange in the loop of the 
\mbox{$\gamma t\bar{t}$} and \mbox{$Z t\bar{t}$} vertex. We give the 
analytic expressions and present numerical results. 

\end{abstract}
\end{titlepage}

\section{Introduction} 
 
Owing to its large mass~\cite{mtop} the top quark decays before its 
polarization is diluted by hadronization effects. This implies 
that one can keep track of its polarization by the distribution  
of its decay products. This property of the top quark  
offers new possibilities of testing existing models. 
New measurable quantities are provided by the  
polarization and the corresponding calculations  
can be performed perturbatively. The large mass of the top quark 
also allows one to probe physics at a high energy scale, where 
new physics might show up.  

In the last years a number of papers~\cite{{CPviol},{dipgl},{we}} 
considered CP violating 
observables in top quark production as tests for new physics. In   
$e^{+} e^{-}$ annihilation these effects are due to the  
weak $\dzs$ and electric $\dgs$ dipole moment form factors 
of the top. In general, the vertices including the CP violating 
form factors are  
\be  
e \, ( {\cal V}_{\gamma}^{t,\bar{t}} )_{\mu}  
&=& e \,  
\left( \frac{2}{3} \gamma_{\mu} 
\mp i \, \frac{\dgs}{m_{t}} {\cal P}_{\mu} \gamma_{5} 
\right) \, , 
\label{photonvertex} 
\\  
\sfrac{g}{2\cos\Theta_{W}} \, ( {\cal V}_{Z}^{t,\bar{t}} )_{\mu}  
&=& 
\sfrac{g}{2\cos\Theta_{W}} \, \left( 
	\gamma_{\mu} ( g_{V} \pm g_{A} \gamma_{5} )  
\mp i \, \frac{\dzs}{m_{t}} {\cal P}_{\mu} \gamma_{5}  
\right) \, ,
\label{Zvertex} 
\ee 
where \mbox{${\cal P}_{\mu} = p_{t\,\mu} - p_{\bar{t}\,\mu}$},  
\mbox{$g_{V} = (1/2) - (4/3) \sin^{2}\Theta_{W}$},
\mbox{$g_{A} = - (1/2)$}, and 
\mbox{$g = e/\sin\Theta_{W}$} with $e$ the electro--magnetic 
coupling constant and $\Theta_{W}$ the 
Weinberg angle. The possibilities of the 
present experimental facility to measure different CP 
violating observables in processes with top quarks
have been studied in~\cite{C.J.}.
As all CP violating observables that can be 
measured depend on these dipole  
moment form factors, precise predictions are necessary. 
 
In the Standard Model (SM) CP violation can  
appear only through the phase in the CKM--matrix.  
The dipole moment form factors $\dgs$ and $\dzs$ for the quarks  
are at least two--loop order effects and hence very small.  
CP violating effects at one--loop level can arise from new  
interactions, and therefore may be larger. The possibility of 
observing CP violating effects in top physics is associated with new 
physics. 
 
In this paper we analyse the supersymmetric (SUSY) contributions 
to $\dgs$ and $\dzs$ of the top quark. In the Minimal  
Supersymmetric Standard Model (MSSM)~\cite{Kane} additional  
complex couplings can be introduced that lead to  
CP violation within one generation only~\cite{Dugan}. If the masses 
of the SUSY particles are not very much higher than the mass  
of the top, one expects SUSY radiative corrections to  
induce larger values for $\dgs$ and $\dzs$. 
 
There is an essential
difference between the dipole moments of the  
electron $d_{e}$ (or neutron $d_{n}$) and the dipole  
moment form factors $\dgs$, $\dzs$ of the top: $d_{e}$  
($d_{n}$) measures the interaction of the stable electron  
(neutron) with almost free photons; $\dgs$ and $\dzs$ are  
form factors that measure the interaction of the short--lived  
top quark with an off--shell $\gamma$ or $Z$.  
The off--shellness of $\gamma$ and $Z$ leads to strong  
enhancements in $\dgs$, $\dzs$ if the particles in the  
loop that couple to $\gamma$, $Z$ are near the threshold  
\mbox{$\sqrt{s} = m_{j} + m_{k}$}, with  
$m_{j}$, $m_{k}$ being the masses of these particles.  
According to the particle content in  
the loop we distinguish the following three contributions:  
\begin{enumerate} 
	\item  
	  gluino contribution $d^{\gamma, Z}_{\tilde{g}}$ with  
	  (\,$\tilde{t}\,\tilde{t}^{*}\,\tilde{g}$\,) in the loop, 
	\item  
	  chargino contribution $d^{\gamma, Z}_{\tilde{\chi}^{+}}$ 
	  with (\,$\tilde{\chi}^{+}\,\tilde{\chi}^{-}\,\tilde{b}$\,) 
	  and (\,$\tilde{b}\,\tilde{b}^{*}\,\tilde{\chi}^{+}$\,) in the loop, 
	\item  
	  neutralino contribution $d^{\gamma, Z}_{\tilde{\chi}^{0}}$ 
	  with (\,$\tilde{\chi}^{0}\,\tilde{\chi}^{0}\,\tilde{t}$\,) 
	  and (\,$\tilde{t}\,\tilde{t}^{*}\,\tilde{\chi}^{0}$\,) in the loop. 
\end{enumerate} 
 
The gluino contribution $d^{\gamma, Z}_{\tilde{g}}$ was considered in  
\cite{{dipgl},{we}}. In this article we consider the chargino and 
neutralino contributions. Although the gluino contribution  
$d^{\gamma, Z}_{\tilde{g}}$ is proportional to $\alpha_{s}$ we 
show that the chargino contribution  
$d^{\gamma, Z}_{\tilde{\chi}^{+}}$, which is  
proportional to $\alpha_{w}$,  
\mbox{($\alpha_{w} = g^{2}/(4\pi)$)}  
can be equally important. This is due to  
threshold enhancements and the large Yukawa couplings: 
\mbox{$Y_{t} = m_{t}/(\sqrt{2} m_{W} \sin\beta)$} and 
\mbox{$Y_{b} = m_{b}/(\sqrt{2} m_{W} \cos\beta)$}.  
In general the neutralino contribution turns out to be smaller. 
However, there are cases where it is important. The mass spectra of 
the SUSY particles are crucial for the understanding of the 
electroweak dipole moment form factors. 
 
In order to obtain a non--zero dipole moment one needs both  
an operator changing the helicity of the top and a mixture of 
the interaction eigenstates of the squarks, charginos, and 
neutralinos, which provide a CP violating phase. In the MSSM the 
CP violating phases for the chargino and neutralino  
contribution are provided by the chargino and neutralino mass 
matrices and the scalar quark mass mixing matrices. The 
calculation requires the diagonalization of all these matrices. 
The chargino mass matrix with complex phases has  
been diagonalized analytically in \cite{Osh1}. In the present paper 
we use the singular value decomposition \cite{SingW} to diagonalize 
the complex neutralino and chargino mass matrices. 
 
The size of the dipole moment form factors $\dgs$, $\dzs$  
depends strongly on the phases of the SUSY parameters. 
There are constraints \cite{{Osh1},{Garisto},{new}} on some phases  
from the measurement of the EDM of the neutron. Usually, 
one concludes~\cite{{Osh1},{Osh2}}  
that either the phases involved in the EDM of the neutron are  
very small or the masses of the first generation of squarks 
are in the TeV range. By using supergravity (SUGRA) with grand 
unification (GUT) there are attempts 
to constrain also the phases entering the dipole moments of the  
top \cite{{Garisto},{new}}. In our analysis we want to be more 
general and we do not make any additional assumptions about GUT 
except for the unification of the gauge couplings. In particular, 
we do not assume unification of the scalar mass parameters and 
the trilinear scalar coupling parameters $A_{q}$  
of the different generations. As the breaking mechanism of SUSY is 
not known, a definite theoretical ansatz about the size of the 
complex phases involved in the weak and electric dipole moment 
form factors of the top quark cannot be given. An unambigous decision 
will be provided by experiment. Therefore, measuring the 
weak and electric dipole moment form factors of the top quark 
might be also a test of the unification assumption involved  
and could give insight into the SUSY breaking mechanism. 
 
In this paper we present a complete analysis of the weak and 
electric dipole moment form factors of the top quark within the 
MSSM with complex parameters. 
We give the analytic expressions and present numerical results.  

The paper is organized as follows. In section~2 we summarize the 
parts of the Lagrangian that are relevant, and fix the notation.  
Generic formulae for contributions to the dipole moment form factors  
are given in section~3. In section~4 we present the chargino 
contribution, in section~5 follows the neutralino contribution.
The numerical analysis is performed in section~6. The conclusions 
are given in section~7. 
 
\section{The SUSY Lagrangian with complex couplings}

In the MSSM with complex phases, $\dgs$ and $\dzs$ are generated in 
one--loop order, irrespectively of generation mixing. 
The gluino contribution to the electroweak dipole moment form 
factors was previously studied in \cite{we}. 
Here we discuss the chargino and neutralino contributions. 
These contributions have to be treated 
separately from the gluino contribution 
not only because different couplings are involved (electroweak 
and strong), but also because they are sensitive to different 
SUSY parameters. The contributions from the different Feynman 
diagrams (Fig.~1a,b) depend in a distinctive way on the
SUSY parameters. 
For the sake of clarity we discuss the chargino and neutralino 
contributions separately.

The parts of the Lagragian that contribute to the 
electric and weak dipole moment form factors of the top are:
\be
 {\mathcal L}_{\gamma\,\tilde{\chi}^{+}_{j}\,\tilde{\chi}^{-}_{k}} 
&=& 
- e \, A_{\mu} \delta_{jk} \,
  \bar{\tilde{\chi}}^{+}_{j} \, \gamma^{\mu} 
  \, \tilde{\chi}^{+}_{k}
\label{LagrangianAchpchm}
\\
 {\mathcal L}_{\gamma \,\tilde{q}_{m}\,\tilde{q}_{n}} 
&=& 
- i e \, Q_{\tilde{q}} A_{\mu} \delta_{mn}
  \, ( \tilde{q}^{*}_{m} \plr^{\mu} \tilde{q}_{n} )
\label{LagrangianAqtqt}
\\
 {\mathcal L}_{Z \,\tilde{\chi}^{+}_{j}\,\tilde{\chi}^{-}_{k}} 
&=& 
  g / ( 2 \cos\Theta_{W} ) \, Z_{\mu} \, 
  \bar{\tilde{\chi}}^{+}_{j} \, \gamma^{\mu}
  ( O^{L}_{jk} P_{L} + O^{R}_{jk} P_{R} ) 
  \, \tilde{\chi}^{+}_{k}
\label{LagrangianZchpchm}
\\
\La_{Z\,\bar{\tilde{\chi}}^{0}_{j}\,\tilde{\chi}^{0}_{k}}
&=&
 g / ( 2 \cos\Theta_{W} )  \,
 Z_{\mu} \, \bar{\tilde{\chi}}^{0}_{j} \, \gamma^{\mu}
 ( O_{jk}^{\prime\prime} P_{L} - O_{jk}^{\prime\prime*} P_{R} )
 \, \tilde{\chi}^{0}_{k}
\label{LagrangianZch0ch0}
\\
 {\mathcal L}_{Z \,\tilde{q}_{m}\,\tilde{q}_{n}} 
&=& 
  i g / ( 2 \cos\Theta_{W} ) \, Z_{\mu} \,
  c^{\tilde{q}}_{mn}
  \, ( \tilde{q}^{*}_{m} \plr^{\mu} \tilde{q}_{n} )
\label{LagrangianZqtqt}
\\
 {\mathcal L}_{\bar{q}\,\tilde{\chi}^{0}_{k}\,\tilde{q}_{m}} 
&=& 
 g \,
 \bar{q} \, 
 ( a_{mk}^{\tilde{q}} P_{R} + b_{mk}^{\tilde{q}} P_{L} ) \, 
 \tilde{\chi}^{0}_{k} \, \tilde{q}_{m}
\label{Lagrangianqch0qt}
\\ 
 {\mathcal L}_{q^{\prime} \,\tilde{\chi}^{+}_{k}\,\tilde{q}_{m}} 
&=& 
  g \, 
  \bar{q}^{\prime} \, 
  ( l_{mk}^{\tilde{q}} P_{R} + k_{mk}^{\tilde{q}} P_{L} ) \, 
  \tilde{\chi}^{+}_{k} \, \tilde{q}_{m}
\label{Lagrangianqschpqt}
\\
 {\mathcal L}_{\bar{q}\,\tilde{g}\,\tilde{q}_{m}} 
&=& 
 - g_{s} \, ( \lambda^{b}_{uv} / \sqrt{2} ) \,
 \bar{q}_{u} \, 
 ( e^{ \frac{i}{2} \varphi_{\tilde{g}}}
   ({\mathcal R}^{\tilde{q}})^{*}_{mL} P_{R} 
 - e^{-\frac{i}{2} \varphi_{\tilde{g}}}
   ({\mathcal R}^{\tilde{q}})^{*}_{mR} P_{L} ) \, 
 \tilde{g}^{b} \, \tilde{q}_{mv}
\label{Lagrangianqgtqt}
\ee
In eq. (\ref{Lagrangianqgtqt}) $\lambda^b$ are the 
Gell--Mann matricies, 
$u,v$ are the color indicies, and $\varphi_{\tilde{g}}$ is the phase 
of the gluino mass term. The couplings 
are defined as in \cite{sabine}:
\be 
O^{R}_{jk} = - ( \cos 2\Theta_{W} \delta_{jk} + U^{*}_{j1} U_{k1} )
\qquad
O^{L}_{jk} = - ( \cos 2\Theta_{W} \delta_{jk} + V_{j1} V^{*}_{k1} )
\ee
\be
O_{jk}^{\prime\prime} = - \sfrac{1}{2}
 [ \cos 2\beta\,
  ( N_{3 j}^{*} N_{3 k} - N_{4 j}^{*} N_{4 k} )
 + \sin 2\beta\,
  ( N_{3 j}^{*} N_{4 k} + N_{4 j}^{*} N_{3 k} )
 ]
= O_{kj}^{\prime\prime*}
\label{Ocoupling}
\ee
\be & 
c^{\tilde{q}}_{mn}
= 2 Q_{\tilde{q}} \sin^{2}\Theta_{W} \delta_{mn}
- 2 \mbox{I}_{3}^{\tilde{q}}
    ( {\mathcal R}^{\tilde{q}} )_{mL} 
    ( {\mathcal R}^{\tilde{q}} )^{*}_{nL}
\ee
\be
a_{mj}^{\tilde{t}} 
= {\mathcal R}^{\tilde{t}*}_{mL}\, f_{Lj}
+ {\mathcal R}^{\tilde{t}*}_{mR}\, h_{Rj}
& \qquad &
b_{mj}^{\tilde{t}} 
= {\mathcal R}^{\tilde{t}*}_{mL}\, h_{Lj}
+ {\mathcal R}^{\tilde{t}*}_{mR}\, f_{Rj}
\ee
\be
h_{Lj} &=& Y_{t} ( \sin\beta\, N_{3 j} - \cos\beta\, N_{4 j} ) 
\qquad
h_{Rj} = Y_{t} ( \sin\beta\, N_{3 j}^{*} - \cos\beta\, N_{4 j}^{*} ) 
       = h_{Lj}^{*}
\nn \\
f_{Lj} &=&
- [ \sfrac{2}{3} \sin 2\,\Theta_{W} \, N_{1 j}^{*} 
  + ( g_{V} - g_{A} ) N_{2 j}^{*} ] / ( \sqrt{2} \cos\Theta_{W} ) 
\label{hcoupling}
\\
f_{Rj} &=& \quad
  [ \sfrac{2}{3} \sin 2\,\Theta_{W} \, N_{1 j} 
  + ( g_{V} + g_{A} ) N_{2 j} ] / ( \sqrt{2} \cos\Theta_{W} ) 
\nn \ee
\be
l_{mj}^{\tilde{b}} 
= - {\mathcal R}^{\tilde{b}*}_{mL}\, U_{j1}
  + {\mathcal R}^{\tilde{b}*}_{mR}\, Y_{b} U_{j2}
& \qquad &
k_{mj}^{\tilde{b}} 
= {\mathcal R}^{\tilde{b}*}_{mL}\, Y_{t} V^{*}_{j2}
\ee
\be
{\mathcal R}^{\tilde{q}} =
    \left(\begin{array}{rr}
      {\mathcal R}^{\tilde{q}}_{1L}
    & {\mathcal R}^{\tilde{q}}_{1R}
  \\  {\mathcal R}^{\tilde{q}}_{2L}
    & {\mathcal R}^{\tilde{q}}_{2R}
    \end{array}\right)
=   \left(\begin{array}{rr}
      e^{ \frac{i}{2} \varphi_{\tilde{q}}} \cos\theta_{\tilde{q}}
    & e^{-\frac{i}{2} \varphi_{\tilde{q}}} \sin\theta_{\tilde{q}}
  \\ -e^{ \frac{i}{2} \varphi_{\tilde{q}}} \sin\theta_{\tilde{q}}
    & e^{-\frac{i}{2} \varphi_{\tilde{q}}} \cos\theta_{\tilde{q}}
    \end{array}\right)
\; , 
\ee
so that ${\tilde{q}_{1} \choose \tilde{q}_{2}} 
= {\mathcal R}^{\tilde{q}} {\tilde{q}_{L} \choose \tilde{q}_{R}}$.
$Q_{\tilde{q}}$ is the charge of $\tilde{q}$ and 
\mbox{$\tan\beta = \sfrac{v_{2}}{v_{1}}$}, where $v_{1}$ and $v_{2}$ 
are the vacuum expectations values of the Higgs fields, $P_{L,R}$ 
are the chirality projection operators.  

The unitary matrices $U$ and $V$ diagonalize the chargino mass 
matrix:
\be
M^{\tilde{\chi}^{+}}_{\alpha\beta} =
\left(
\begin{array}{cc}
  M                        & m_{W} \sqrt{2} \sin\beta  \\
  m_{W} \sqrt{2} \cos\beta & \mu      
\end{array}
\right)
\; , \qquad
U^{*}_{j\alpha} M^{\tilde{\chi}^{+}}_{\alpha\beta} V^{*}_{k\beta} 
= m_{\tilde{\chi}_{j}^{+}} \delta_{jk}
\; . \qquad
\label{charginomassmatrix}
\ee
For the diagonalization of the chargino mass matrix 
one has to use the singular value 
decomposition~\cite{SingW}. The explicit procedure to obtain 
$U$ and $V$ is given in Appendix~A.

$N_{\alpha k}$ is the unitary matrix which 
makes the complex symmetric neutralino mass matrix diagonal 
with positive diagonal elements~\cite{BilPet}: 
\be
N_{\alpha j} M^{\tilde{\chi}^{0}}_{\alpha\beta} N_{\beta k} 
= m_{\tilde{\chi}_{j}^{0}} \delta_{jk}
\; .
\ee
In the basis 
\be
\psi_{\alpha} = 
\{ -i\tilde{\gamma}, -i\tilde{Z}, \tilde{H}_{1}, \tilde{H}_{2} \} 
\quad \mbox{\small with} \;
\left\{
\begin{array}{ccc}
 \tilde{H}_{1} &=& \tilde{H}_{1}^{0} \cos\beta 
    - \tilde{H}_{2}^{0} \sin\beta  \\ \tilde{H}_{2} &=& 
\tilde{H}_{1}^{0} \sin\beta 
     + \tilde{H}_{2}^{0} \cos\beta 
\end{array}\right.
\ee
used in \cite{Kane}, the complex symmetric neutralino 
mass matrix has the form 
\be
M^{\tilde{\chi}^{0}}_{\alpha\beta} =
\left(
\begin{array}{cccc}
m_{\tilde{\gamma}} &        m_{az} &      0 &   0 \\ 
            m_{az} & m_{\tilde{z}} &  m_{Z} &   0 \\
0 & m_{Z} &   \mu \sin 2\beta & - \mu \cos 2\beta \\ 
0 &     0 & - \mu \cos 2\beta & - \mu \sin 2\beta
\end{array}
\right)
\label{neutralinomassmatrix}
\ee
where 
\be
m_{\tilde{\gamma}} 
&=& M \sin^{2}\Theta_{W}
 + M^{\prime} \cos^{2}\Theta_{W}
\; ,
\nn \\
m_{\tilde{z}} 
&=& M \cos^{2}\Theta_{W}
 + M^{\prime} \sin^{2}\Theta_{W}
\; ,
\\
m_{az} 
&=& \sin\Theta_{W} \cos\Theta_{W} ( M - M^{\prime} ) 
\; .
\nn \ee
Here $M$ and $M'$ are the SU(2) and U(1) gaugino masses, 
$\mu = |\mu| e^{i\varphi_{\mu}}$ is the mass parameter in front of the 
Higgs superfields in the Lagrangian. 
The procedure for obtaining $N_{\alpha k}$ is described in 
Appendix~B. Again the singular value decomposition has to be used. 

\section{General formula}

Technically there 
are two types of one--loop SUSY diagrams that contribute 
to the electric and weak dipole moments of the top quark, one with 
two scalars and one fermion	in the loop as shown in \mbox{Fig.~1a},
and the other with two fermions and one scalar in the loop, 
\mbox{Fig.~1b}. 
Here we give generic formulae for the dipole moment form factors 
of the top induced by these graphs. The different combinations of 
Passarino--Veltman functions give different types of 
threshold enhancements. 

We use a generic Lagrangian of the form 
\be
\La
&=&
  g_{1}  \, V_{\mu} \, \bar{f}_{j} \, \gamma^{\mu}
  ( O_{jk}^{L} P_{L} + O_{jk}^{R} P_{R} ) \, f_{k}
+ i g_{1} \, V_{\mu} \, \Gamma_{mn}
  \, ( s^{*}_{m} \plr^{\mu} s_{n} )
\nn \\
&+& 
 g \, \bar{f} \, ( a_{mk} P_{R} + b_{mk} P_{L} ) \, f_{k} \, s_{m}
+ h.c.
\ee
where $f_{i}$ stands for a fermion field, $s_{m}$ for a scalar 
field, and $V_{\mu}$ for a neutral vector boson field. The couplings 
$\Gamma_{mn}$ and $O_{jk}^{L,R}$ are hermitian. The couplings 
$a_{mk}$ and $b_{mk}$ are complex numbers. Fig.~1a then gives 
the following contribution:
\be
d_{ffs} &=& \alpha_{w} / ( 8 \pi ) \times ( f_{1} + f_{2} )
\\
f_{1} &=& - 2 \sum_{m,j,k} m_{j} C_{1}^{m,jk} \,
  \mIm[ a_{mj} O^{L}_{jk} b_{mk}^{*}
      - b_{mj} O^{R}_{jk} a_{mk}^{*}
      ]
\nn \\
f_{2} &=& m_{t} \sum_{m,j,k}
  ( C_{1}^{m,jk} + C_{11}^{m,jk} - C_{2}^{m,jk} - C_{22}^{m,jk} )
\nn \\ & & \hspace{2cm} \times
  \mIm
    [ a_{mj} O^{R}_{jk} a_{mk}^{*}
    - b_{mj} O^{L}_{jk} b_{mk}^{*}
    ]
\nn \ee
with \mbox{$g^{2} = 4\pi\alpha_{w}$}.
Fig~1b leads to 
\be
d_{ssf} &=& \alpha_{w} / ( 8 \pi ) \times ( s_{1} + s_{2} )
\\
s_{1} &=&
- 2 \sum_{k,m,n}
  m_{k} ( C_{0}^{k,mn} + C_{1}^{k,mn} + C_{2}^{k,mn} ) \,
  \mIm [ a_{mk} \Gamma_{mn} b_{nk}^{*} ]
\nn \\
s_{2} &=& 
- m_{t} \sum_{k,m,n}
  ( C_{1}^{k,mn} + C_{11}^{k,mn} - C_{2}^{k,mn} - C_{22}^{k,mn} )
\nn \\ & & \hspace{2cm} \times
  \mIm
  [ a_{mk} \Gamma_{mn} a_{nk}^{*}
  - b_{mk} \Gamma_{mn} b_{nk}^{*}
  ]
\nn \ee
For the Passarino--Veltman three point functions~\cite{PaVe} 
$C_{0}$, $C_{i}$, and $C_{ii}$ ($i = 1,2$) we follow the convention 
of \cite{Denner}. They are defined in Appendix~C. Note that there are 
only three types of contributions: $f_{2}$ and $s_{2}$ have the same 
functional form. 

\section{Chargino contribution to $\dgs$ and $\dzs$} 
 
Here we show the explicit dependence of the chargino contribution  
on the gaugino and higgsino couplings, as well as the dependence on  
the squark mixing angle and phase. For the  
electric dipole moment form factor we have:  
\be 
\frac{d^{\gamma}_{\tilde{\chi}^{+}}}{m_{t}}
&=& - \frac{\alpha_{em} \, Y_{t}}{4\pi\sin^2\Theta_{W}} \,
\sum_{m,i = 1}^{2} \hspace{-2mm} m_{\tilde{\chi}^{+}_{i}}
( C_{1}^{m,ii}
- \sfrac{1}{6} ( C_{0}^{i,mm} + C_{1}^{i,mm} + C_{2}^{i,mm} ) )
\times
\\ & & \qquad
\Bigl[
    ( 1 - (-)^{m} \cos 2\theta_{\tilde{b}} )
    \, \mIm [ U_{i1} V_{i2} ]
  + Y_{b} (-)^{m} \sin 2\theta_{\tilde{b}}
    \, \mIm [ U_{i2} V_{i2} e^{i\varphi_{\tilde{b}}} ]
\Bigr]
\label{dGammaChargino}
\nn\ee 
The chargino contribution $d^{Z}_{\tilde{\chi}^{+}}$ to the weak 
dipole moment form factor of the top quark is: 
\be 
\frac{d^{Z}_{\tilde{\chi}^{+}}}{m_{t}}
&=& \frac{\alpha_{em} \,}{8\pi\sin^2\Theta_{W}} \,
    ( f_{1} + f_{2} + s_{1} + s_{2} )
\ee
\be 
f_{1}
&=& - Y_{t} \sum_{m,j,k = 1}^{2} \hspace{-1mm} 
m_{\tilde{\chi}^{+}_{j}} C_{1}^{m,jk}
\times 
\Bigl[
  ( \delta_{jk} ( 1 + 2 \cos 2\Theta_{W} ) 
  + |U_{k1}|^{2} - |V_{k2}|^{2} ) \cdot
\\ & & \quad
  ( ( 1 - (-)^{m} \cos 2\theta_{\tilde{b}} )
    \, \mIm [ U_{j1} V_{j2} ]
  + Y_{b} (-)^{m} \sin 2\theta_{\tilde{b}}
    \, \mIm [ U_{j2} V_{j2} e^{i\varphi_{\tilde{b}}} ]
  )
\nn\\ & & \quad
- Y_{b} ( 1 - \delta_{jk} ) (-)^{m} \sin 2\theta_{\tilde{b}}
    \, \mIm [ U_{j2} V_{j2} e^{i\varphi_{\tilde{b}}} ]
\Bigr] 
\nn\\
f_{2}
&=& 
- m_{t} Y_{b} \sin 2\theta_{\tilde{b}}
  \mIm [ U_{11}^{*} U_{12}^{\,} e^{i\varphi_{\tilde{b}}} ]
\sum_{m=1}^{2} (-)^{m}
  ( C_{1}^{m,12} + C_{11}^{m,12} - C_{2}^{m,12} - C_{22}^{m,12} )
\\
s_{1}
&=& - Y_{t}
\sum_{k,m,n=1}^{2} \hspace{-1mm}
m_{\tilde{\chi}^{+}_{k}}
  ( C_{0}^{k,mn} + C_{1}^{k,mn} + C_{2}^{k,mn} )
\\ & & \qquad
\times 
  ( \sfrac{2}{3} \sin^{2} \Theta_{W} \delta_{mn}
  + \sfrac{1}{2} ( 1 - (-)^{n} \cos 2\theta_{\tilde{b}} )
  ) 
\nn\\ & & \qquad \cdot
\Bigl[
  ( ( 1 - (-)^{m} \cos 2\theta_{\tilde{b}} )
    \, \mIm [ U_{k1} V_{k2} ] 
  + Y_{b} (-)^{m} \sin 2\theta_{\tilde{b}}
    \, \mIm [ U_{k2} V_{k2} e^{i\varphi_{\tilde{b}}} ]
\Bigr] 
\nn\\
s_{2}
&=& 
- m_{t} Y_{b} \sin 2\theta_{\tilde{b}}
\sum_{k=1}^{2} 
  ( C_{1}^{k,12} + C_{11}^{k,12} - C_{2}^{k,12} - C_{22}^{k,12} )
\times
  \mIm [ U_{k1}^{*} U_{k2}^{\,} e^{i\varphi_{\tilde{b}}} ]
\label{dZChargino}
\ee 
We have included the terms proportional to the bottom Yukawa 
coupling $Y_{b}$ which are important for large $\tan\beta$. For small 
values of $\tan\beta$ the terms proportional to $Y_{t}$ dominate. The 
exchange of the lighter scalar bottom (see Fig.~1a) gives the leading 
contribution ($\cos 2\theta_{\tilde{b}} \approx 1$). Note that 
there are no terms proportional to $Y_{t}^{2}$ or $Y_{b}^{2}$. 

\section{Neutralino contribution to $\dgs$ and $\dzs$} 
 
As  neutralinos do not couple to the photon, 
$d^{\gamma}_{\tilde{\chi}^{0}}$ receives a non--zero 
contribution only 
from the diagram with $\tilde{t}\,\tilde{t}^{*}\,\tilde{\chi}^{0}$ 
exchanged in the loop. We have: 
\be 
\frac{d^{\gamma}_{\tilde{\chi}^{0}}}{m_{t}} 
&=& 
  \frac{\alpha_{em}}{12\pi\sin^2\Theta_{W}} \,
  \sum_{k=1}^{4} \sum_{m=1}^{2} 
 m_{\tilde{\chi}^{+}_{k}}
  ( C_{0}^{k,mm} + C_{1}^{k,mm} + C_{2}^{k,mm} )
\\ & & \quad
\times
\Bigl[ 
  (-)^{m} \sin 2\theta_{\tilde{t}} 
  \mIm [ ( h_{Lk}^{2} - f_{Lk} f_{Rk}^{*} )
         e^{-i \varphi_{\tilde{t}}} ]
\nn \\ & & \qquad 
- ( 1 - (-)^{m} \cos 2\theta_{\tilde{t}} ) \mIm [ h_{Lk} f_{Lk}^{*} ] 
- ( 1 + (-)^{m} \cos 2\theta_{\tilde{t}} ) \mIm [ h_{Lk} f_{Rk} ] 
\Bigr] 
\nn \ee 
The neutralino contribution to $\dzs$ is  
\be 
\frac{d^{Z}_{\tilde{\chi}^{0}}}{m_{t}} 
&=& \frac{\alpha_{em}}{8\pi\sin^2\Theta_{W}} \,
    ( 2 f_{1} + 2 f_{2} + s_{1} + s_{2} )
\label{dZsNeutralino}
\ee
\be
f_{1} &=& 
\sfrac{1}{2} \sum_{j,k=1}^{4} \sum_{m=1}^{2}
 m_{\tilde{\chi}^{0}_{j}} C_{1}^{m,jk}  \times
\Bigl[
  (-)^{m} \sin 2\theta_{\tilde{t}} \,
  \mIm [ O_{jk}^{\prime\prime}
         ( f_{Lj} f_{Rk}^{*} - f_{Lk} f_{Rj}^{*} )
         e^{-i \varphi_{\tilde{t}}} ]
\\ & & \qquad \qquad
- ( 1 + (-)^{m} \cos 2\theta_{\tilde{t}} )
  \mIm [ O_{jk}^{\prime\prime}
         ( h_{Lj}^{*} f_{Rk}^{*} - h_{Lk}^{*} f_{Rj}^{*} ) ]
\nn \\ & & \qquad \qquad
- ( 1 - (-)^{m} \cos 2\theta_{\tilde{t}} )
  \mIm [ O_{jk}^{\prime\prime}
         ( h_{Lj} f_{Lk}^{*} - h_{Lk} f_{Lj}^{*} ) ]
\Bigr] 
\label{dZNeuf1}
\nn \\
f_{2} &=& 
\sfrac{1}{2} m_{t} \sum_{j<k}^{4} \sum_{m=1}^{2} 
  ( C_{1}^{m,jk} + C_{11}^{m,jk} - C_{2}^{m,jk} - C_{22}^{m,jk} )
\\ & & \qquad
\times
\Bigl[ \,
  2 (-)^{m} \cos 2\theta_{\tilde{t}} \,
  \mIm [ h_{Lj} O_{jk}^{\prime\prime} h_{Lk}^{*} ]
\nn \\ & & \qquad \quad
- 2 (-)^{m} \sin 2\theta_{\tilde{t}} \,
  \mIm [ ( f_{Lj}^{*} - f_{Rj} ) O_{jk}^{\prime\prime} h_{Lk}^{*}
         e^{i \varphi_{\tilde{t}}} ]
\nn \\ & & \qquad \quad
+ ( 1 - (-)^{m} \cos 2\theta_{\tilde{t}} )
  \mIm [ f_{Lj}^{*} O_{jk}^{\prime\prime} f_{Lk} ]
\nn \\ & & \qquad \quad
+ ( 1 + (-)^{m} \cos 2\theta_{\tilde{t}} )
  \mIm [ f_{Rj} O_{jk}^{\prime\prime} f_{Rk}^{*} ]
\Bigr] 
\nn\\
s_{1}
&=& 
\sfrac{1}{2} \sum_{k=1}^{4} \sum_{m,n=1}^{2} 
m_{\tilde{\chi}^{+}_{k}}
  ( C_{0}^{k,mn} + C_{1}^{k,mn} + C_{2}^{k,mn} )
\\ & & \quad
\times 
\Bigl[ \, (-)^{m} 
  \Bigl(
    \sfrac{8}{3} \sin^2\Theta_{W} \delta_{mn}
  - ( 1 - (-)^{n} \cos 2\theta_{\tilde{t}} )
  \Bigr)
  \mIm [ ( h_{Lk}^{2} - f_{Lk} f_{Rk}^{*} )
         e^{-i \varphi_{\tilde{t}}} ]
\nn \\ & & \qquad 
- \Bigl(
    \sfrac{8}{3} \sin^2\Theta_{W} \delta_{mn}
    ( 1 + (-)^{m} \cos 2\theta_{\tilde{t}} )
  - (-)^{m+n} \sin^{2} 2\theta_{\tilde{t}}
  \Bigr)
  \mIm [ h_{Lk} f_{Rk} ] 
\nn \\ & & \qquad 
- \Bigl(
    \sfrac{8}{3} \sin^2\Theta_{W} \delta_{mn}
  - ( 1 - (-)^{n} \cos 2\theta_{\tilde{t}} )
  \Bigr) 
  ( 1 - (-)^{m} \cos 2\theta_{\tilde{t}} )
  \mIm [ h_{Lk} f_{Lk}^{*} ] 
\Bigr] 
\nn\\
s_{2}
&=& 
m_{t} \sin 2\theta_{\tilde{t}}
\sum_{k=1}^{4}
  ( C_{1}^{k,12} + C_{11}^{k,12} - C_{2}^{k,12} - C_{22}^{k,12} )
  \mIm [ h_{Lk}^{*} ( f_{Lk}^{*} - f_{Rk} )
         e^{i\varphi_{\tilde{t}}} ]
\ee 
$f_{Lj}$, $f_{Rk}$ are gaugino 
couplings and $h_{Lj}$ are higgsino couplings that contain the large 
Yukawa coupling $Y_{t}$. 
Notice that the factor of $2$ in front of $f_{1}$ and 
$f_{2}$ in eq.(\ref{dZsNeutralino}) is due to the Majorana nature 
of the neutralinos. 

\section{Numerical results}

In this section we give numerical results for the real and 
imaginary parts of $\dgs$ and $\dzs$. 
Quite generally they depend on the parameters 
$M^{\prime}$, $M$, $|\mu|$, $\tan\beta$, $m_{\tilde{t}_{k}}$,
$m_{\tilde{b}_{k}}$, $\cos\theta_{\tilde{t}}$, 
$\cos\theta_{\tilde{b}}$ and the phases $\varphi_{\mu}$, 
$\varphi_{\tilde{t}}$, $\varphi_{\tilde{b}}$,  and
$\varphi_{\tilde{g}}$.  
The GUT relations
\be
m_{\tilde{g}} &=& (\alpha_{s}/\alpha_{2}) M \approx 3 M
\\ 
M^{\prime} &=& \sfrac{5}{3} \tan^{2}\Theta_{W} M
\ee
imply that the gaugino mass parameters have the same phase. This phase 
can be removed by a R--transformation~\cite{Dugan}. 

We take \mbox{$m_{W} = 80$~GeV}, \mbox{$m_{t} = 175$~GeV}, 
\mbox{$m_{b} = 5$~GeV}, \mbox{$\sqrt{s} = 500$~GeV}, 
\mbox{$\alpha_{s} = 0.1$}, and \mbox{$\alpha_{em} = \sfrac{1}{123}$}.
In order not to vary too many parameters we choose a reference set of
parameter values given in the following table:
\begin{center}
\begin{tabular}{|rcrl|rcrl|rcrl|}
  \hline
  $M$ &=& 230, 360 & \hspace{-4mm} GeV &
    $m_{\tilde{t}_{1}}$ &=& 150 & \hspace{-4mm} GeV &
      $m_{\tilde{b}_{1}}$ &=& 270 & \hspace{-4mm} GeV \\
  $|\mu|$ &=&      250 & \hspace{-4mm} GeV &
    $m_{\tilde{t}_{2}}$ &=& 400 & \hspace{-4mm} GeV &
      $m_{\tilde{b}_{2}}$ &=& 280 & \hspace{-4mm} GeV \\
  $\tan\beta$ &=& 2 & &
    $\theta_{\tilde{t}}$ &=& $\frac{\pi}{9}$ & &
      $\theta_{\tilde{b}}$ &=& $\frac{\pi}{36}$ & \\
  $\varphi_{\mu}$ &=& $\frac{4 \pi}{3}$ & &
    $\varphi_{\tilde{t}}$ &=& $\frac{\pi}{6}$ & &
      $\varphi_{\tilde{b}}$ &=& $\frac{\pi}{3}$ & \\
  \hline
\end{tabular}
\end{center}
For $\tan\beta = 2$ the terms proportional to $Y_{b}$ are
strongly suppressed, and as one can easily verify from the explicit
expressions (\ref{dGammaChargino})--(\ref{dZChargino})
for $d^{\gamma,Z}_{\tilde{\chi}^{+}}$, the result will be
independent on the phase $\varphi_{\tilde{b}}$.
Thus $\dgs$ and $\dzs$ will depend on two phases: $\varphi_{\mu}$
and $\varphi_{\tilde{t}}$. 

Notice that the dipole moment form factors $\dgs$ and $\dzs$ depend 
on $\varphi_{\mu}$ not only through the mixing matrices $U$, $V$, 
and $N$ in the couplings, but also 
through the chargino and neutralino mass spectra. 
In Fig.~2a and b we show the dependence of the chargino and 
neutralino masses on the phase $\varphi_{\mu}$ for $M = 230$~GeV and 
the other parameters as in the reference set. As can be seen the 
values of the masses vary by about 40 percent in the whole 
\mbox{$\cos\varphi_{\mu}$} region. 

Next we study the dependence of $\dgs$ and $\dzs$ on the parameters 
$M$ and $|\mu|$ which control the chargino and neutralino 
masses and couplings. 
In Fig.~3a we show the $|\mu|$ dependence of the chargino 
contributions to \mbox{$\mIm \dgs$} and \mbox{$\mIm \dzs$} keeping 
the other parameters fixed at the reference values.
 
\mbox{$\mIm \dgs$} and \mbox{$\mIm \dzs$} are
determined by the absorptive parts of the amplitudes. Therefore they 
vanish when no real production of charginos is possible, i.e. 
\mbox{$\sqrt{s} \le 2 m_{\tilde{\chi}^{+}_{1}}$} as can be seen 
from the dotted curve of Fig.~3a. 
Local maxima occur near the thresholds of chargino pair production 
and they get bigger if the gaugino and higgsino component of the 
chargino are approximately equal. The 
$\tilde{\chi}^{+}_{1} \tilde{\chi}^{-}_{1}$--contribution 
has always the opposite sign of the
$\tilde{\chi}^{+}_{2} \tilde{\chi}^{-}_{2}$--contribution 
because of the coupling \mbox{$\mIm ( U_{i1} V_{i2} )$} 
(see eq.(\ref{dGammaChargino})). 

In Fig.~4a we show the neutralino contribution to \mbox{$\mIm \dgs$} 
and \mbox{$\mIm \dzs$}. \mbox{$\mIm d^{\gamma}_{\tilde{\chi}^{0}}$} 
is one order of magnitude 
smaller than the chargino contribution because the photon does not 
couple to the neutralinos and only Fig.~1b contributes. 
The neutralino contribution \mbox{$\mIm d^{Z}_{\tilde{\chi}^{0}}$} 
is smaller than the chargino 
contribution because the couplings are smaller. It 
shows the same qualitative behaviour as the chargino contributions,
but it is more complicated because of the richer particle spectrum. 
Note that the two neutralinos in Fig.~1a have to be different.   

In Fig.~3b we show the chargino contribution to \mbox{$\eRe \dgs$} 
and \mbox{$\eRe \dzs$} as a function of $|\mu|$ for 
\mbox{$M =$ 230, 360 GeV} and the other parameters as given 
in the reference set. The behaviour of 
\mbox{$\eRe d^{\gamma}_{\tilde{\chi}^{+}}$} and 
\mbox{$\eRe d^{Z}_{\tilde{\chi}^{+}}$} can be understood by 
the dispersion relations by which \mbox{$\eRe \dgs$} and 
\mbox{$\eRe \dzs$} 
are related to the absorptive parts of $\dgs$ and $\dzs$.
In Fig.~4b we show the neutralino contribution 
\mbox{$\eRe d^{Z}_{\tilde{\chi}^{0}}$}. As one can see it has an 
analogous behaviour as the chargino contribution (Fig.~3b), but 
it is smaller. 

In Fig.~5a we show the dependence of the chargino contribution 
\mbox{$\mIm d^{\gamma}_{\tilde{\chi}^{+}}$} and 
\mbox{$\mIm d^{Z}_{\tilde{\chi}^{+}}$} on the phase $\varphi_{\mu}$. 
Note that for \mbox{$M = 360$ GeV}
\mbox{$\mIm d^{\gamma}_{\tilde{\chi}^{+}}$} and 
\mbox{$\mIm d^{Z}_{\tilde{\chi}^{+}}$} vanish in the intervall
\mbox{$0.82\pi < |\varphi_{\mu}| < 1.18\pi$}. 
This reflects the dependence of the chargino masses on the 
phase $\varphi_{\mu}$ (Fig.~2a):
$m_{\tilde{\chi}^{+}_{1}}$ decreases with $\cos\varphi_{\mu}$, and 
for \mbox{$0.82\pi < |\varphi_{\mu}| < 1.18\pi$} we are below the 
threshold: \mbox{$\sqrt{s} \leq 2 m_{\tilde{\chi}^{+}_{1}}$}. The 
maxima of \mbox{$\mIm \dgs$}, \mbox{$\mIm \dzs$} for the parameters 
used appear at \mbox{$\varphi_{\mu} = 0.65 \pi$}, and 
\mbox{$\varphi_{\mu} = 0.40 \pi$}, and not at 
\mbox{$\varphi_{\mu} = 0.5 \pi$} as one would naively expect. 

In Fig.~5c one can clearly see the difference between the two types 
of diagrams Fig.~1a and Fig.~1b: as the photon couples only to the 
stops, \mbox{$\mIm d^{\gamma}_{\tilde{\chi}^{0}}$} has a smooth 
dependence on $\varphi_{\mu}$, 
whereas \mbox{$\mIm d^{Z}_{\tilde{\chi}^{0}}$} depends on 
$\varphi_{\mu}$ through the 
masses of the neutralinos. The thresholds of neutralino 
production are clearly visible at \mbox{$\varphi_{\mu} = 0.6 \pi$} 
and \mbox{$\varphi_{\mu} = 0.77 \pi$}. Fig.~5d and Fig.~5c are 
connected via the dispersion relations for $\dgs$ and $\dzs$.

The chargino contribution exhibits only a very small dependence on 
$\theta_{\tilde{b}}$ and $\varphi_{\tilde{b}}$ for the parameter 
region chosen. For \mbox{$\mIm \dgs$} it turns out that one can 
neglect the neutralino contributions, so the dependence on the 
mixing angle $\theta_{\tilde{t}}$ and the phase $\varphi_{\tilde{t}}$ 
is irrelevant. However for \mbox{$\mIm \dzs$} the neutralino 
contributions are quite big. The main contributions come from the 
second and the third part of eq.(\ref{dZNeuf1}) which are proportional 
to $\cos 2\theta_{\tilde{t}}$. As $\varphi_{\tilde{t}}$ does not 
appear in these contributions, the dependence of 
$d^{Z}_{\tilde{\chi}^{0}}$ on $\varphi_{\tilde{t}}$ is small. 

There is a smooth dependence on $m_{\tilde{b}}$: 
\mbox{$\mIm d^{\gamma}_{\tilde{\chi}^{+}}$} and 
\mbox{$\mIm d^{Z}_{\tilde{\chi}^{+}}$} increase from $-0.0014$ 
at \mbox{$m_{\tilde{b}_{1}}=100$~GeV} to $-0.0004$ at 
\mbox{$m_{\tilde{b}_{1}}=400$~GeV}. The dependence of 
$d^{\gamma}_{\tilde{g}}$ on 
$m_{\tilde{t}}$ is already shown in \cite{we}. 
$d^{Z}_{\tilde{\chi}^{0}}$ exhibits 
a smooth dependence on $m_{\tilde{t}_{1}}$: 
\mbox{$\mIm d^{Z}_{\tilde{\chi}^{0}}$} decreases from $0.0007$ 
at \mbox{$m_{\tilde{t}_{1}}=100$~GeV} to $0.0003$ at 
\mbox{$m_{\tilde{t}_{1}}=250$~GeV}. 

In Fig.~6a we show $\dgs$ and $\dzs$ as functions of $\sqrt{s}$ where 
all contributions (gluino, chargino, and neutralino) are summed up. 
The threshold effects can be seen very clearly.
There is a big enhancement in $\dgs$ and $\dzs$ because the 
threshold for $\tilde{\chi}^{+}_{1}\tilde{\chi}^{-}_{1}$ production 
is reached at \mbox{$\sqrt{s}=420$~GeV} for 
\mbox{$m_{\tilde{\chi}^{+}_{1}} = 210$~GeV}. At 
\mbox{$\sqrt{s}=590$~GeV} $\tilde{\chi}^{+}_{2}\tilde{\chi}^{-}_{2}$ 
production becomes possible and again there is a big contribution but 
with a different sign. The additional thresholds in $\dzs$ are due to 
the neutralino contributions. Fig.~6b can again be understood via 
dispersion relations: each spike corresponds to the opening of a new 
production channel.

\section{Summary and Conclusions}

We have calculated all contributions to the electric and weak dipole 
moment form factors of the top quark, $\dgs$ and $\dzs$, within the 
MSSM with complex SUSY parameters $\mu$, $A_{t}$, $A_{b}$, and 
$m_{\tilde{g}}$. These form factors can be measured by the 
reaction \mbox{$e^{+}e^{-} \to t\,\bar{t}$} with \mbox{$t \to b W$}. 
They are different from zero only if CP is violated. 
After the diagonalization of the complex chargino, neutralino, 
stop, and sbottom mass matrices, the CP violating phases of the 
couplings are related to the phases of $\mu$, $A_{t}$, $A_{b}$, 
and $m_{\tilde{g}}$.

We find that the chargino contribution is even larger than the gluino 
contribution for \mbox{$m_{\tilde{g}} \gsim 500$~GeV} due to the 
Yukawa couplings. The neutralino exchange plays a less important 
role, but must not be neglected. The dependence of $\dgs$ and $\dzs$ 
on the SUSY parameters and on the energy is very characteristic. 
There are enhancements whenever the particles in the loop that 
couple to $\gamma$, $Z$ reach a threshold. Therefore by measuring 
the dipole moment form factors by suitable asymmetries one can get
information about the SUSY parameters. 

It is important to point out that we have performed our analysis 
within the general framework of the MSSM with complex parameters. In 
particular, we have not used universal SUSY parameters at the GUT 
scale. 
The numerical values of the form factors can reach about $10^{-3}$, 
which in general results in a measurable asymmetry of this size.  

\section*{Acknowledgements}

We thank Helmut Eberl for his constructive assisitance in the 
evaluation of the loop integrals and Sabine Kraml for the discussions 
about the couplings in the Lagrangian. 
We also thank Stefano Rigolin for his helpful correspondence 
regarding the numerical calculations. 
E.C.'s work has been 
supported by the Bulgarian National Science Foundation, Grant Ph--510. 
This work was also supported by the 'Fonds zur F\"orderung der 
wissenschaftlichen Forschung' of Austria, project no. P10843--PHY. 

\newpage 

\section*{Appendix A}

Here we show how to obtain the unitary matrices 
$U_{\alpha k}$ and $V_{\alpha k}$ from the 
complex chargino mass matrix $M_{\alpha\beta}$ 
eq.(\ref{charginomassmatrix}). 
For arbitrary $M_{\alpha\beta}$ we have 
\be
  M_{\alpha\beta} U_{\alpha j}^{*} V_{\beta k}^{*} 
= \delta_{jk} m_{k}
\quad \mbox{\small and} \quad
M_{\alpha\beta} = U_{\alpha k} V_{\beta k} m_{k}
\label{diagonalization}
\ee
with unitary 
\be
U_{\alpha j}^{*} U_{\alpha k} = \delta_{jk}
\quad
U_{\alpha k} U_{\beta k}^{*} = \delta_{\alpha\beta}
\quad
V_{\alpha j} V_{\alpha k}^{*} = \delta_{jk}
\quad
V_{\alpha k}^{*} V_{\beta k} = \delta_{\alpha\beta}
\quad \mbox{\small and} \;
m_{k} \ge 0
\; .
\label{unitarity}
\ee
Obtaining $V_{\alpha k}$ from 
\be
  ( M_{\alpha\beta} U_{\alpha j}^{*} V_{\beta k}^{*} )^{*}
  M_{\alpha'\beta'} U_{\alpha' j}^{*} V_{\beta' l}^{*}
= M_{\alpha\beta}^{*} M_{\alpha'\gamma} \delta_{\alpha'\alpha}
  V_{\beta k} V_{\gamma l}^{*}
= ( \delta_{jk} m_{k} )^{*} \delta_{jl} m_{l}
= \delta_{kl} m_{k}^{2}
\ee
one sees
\be
  \delta_{kl} m_{k}
= U_{\alpha k}^{*} M_{\alpha\beta} V_{\beta l}^{*} 
= ( 1 / m_{k} ) \delta_{kl} m_{k}^{2}
= ( 1 / m_{k} ) M_{\alpha\gamma}^{*} V_{\gamma k}
  M_{\alpha\beta} V_{\beta l}^{*} 
\ee
that 
\be
  U_{\alpha k}^{*} 
= ( 1 / m_{k} ) M_{\alpha\gamma}^{*} V_{\gamma k}
\qquad \mbox{\small or} \quad
  U_{\alpha k}
= M_{\alpha\gamma} V_{\gamma k}^{*} / m_{k}
\; .
\ee

\section*{Appendix B}

Here we show how to obtain the unitary matrix $N_{\alpha k}$ from the 
complex symmetric neutralino mass matrix $M_{\alpha\beta}$ 
eq.(\ref{neutralinomassmatrix}) like in \cite{BilPet}. 

For arbitrary $M_{\alpha\beta}$ we have eq.(\ref{diagonalization}) 
and (\ref{unitarity}) to obtain $u_{\alpha j}$ and $v_{\beta k}$. 
Because $M_{\alpha\beta} = M_{\beta\alpha}$ is symmetric 
$ u_{\alpha k} v_{\beta k} m_{k} 
= u_{\beta k} v_{\alpha k} m_{k}$ or
\be 
  u_{\alpha k} v_{\beta k} m_{k} 
  \cdot v_{\alpha j}^{*} v_{\beta l}^{*} 
= v_{\alpha j} u_{\alpha l}^{*} m_{l}
& \quad &
  u_{\beta k} v_{\alpha k} m_{k} 
  \cdot v_{\alpha j}^{*} v_{\beta l}^{*} 
= u_{\beta j} v_{\beta l}^{*} m_{j}
\nn \\ \mbox{\small so}\quad 
  v_{\alpha j}^{*} u_{\alpha l}
= v_{\alpha l}^{*} u_{\alpha j}
= 0
& \mbox{\small for} & \; m_{j} \neq m_{l}
\ee
so 
\be & & 
s^{2}_{jk} 
:= v_{\alpha j}^{*} u_{\alpha k}
= e^{2 i \alpha_{k}} \delta_{jk}
\; \mbox{\small with} \;
0 \le \alpha_{k} < \pi
\qquad \mbox{\small since} \quad
s^{2*}_{jk} s^{2}_{kl} = \delta_{jl}
\; ,
\nn \\ & &
  v_{\beta j} s^{2}_{jk}
= v_{\beta j} v_{\alpha j}^{*} u_{\alpha k}
= u_{\beta k}
\; .
\ee 
With the definitions $s^{1}_{jk} := e^{i \alpha_{k}} \delta_{jk}$
and $v_{\alpha k} = N_{\alpha j} s^{1}_{jk}
 = N_{\alpha k} e^{i \alpha_{k}}$ follows 
\be & &
  u_{\alpha k} 
= v_{\alpha j} s^{2}_{jk}
= N_{\alpha j}^{*} e^{- i \alpha_{j}} e^{2 i \alpha_{k}} \delta_{jk}
= N_{\alpha k}^{*} e^{i \alpha_{k}}
\\ & &
  M_{\alpha\beta} 
= u_{\alpha k} v_{\beta k} m_{k}
= N_{\alpha k}^{*} e^{i \alpha_{k}}
  N_{\beta k}^{*} e^{-i \alpha_{k}}
  m_{k}
= N_{\alpha k}^{*} N_{\beta k}^{*} m_{k}
\; .
\ee

\section*{Appendix C}

Here we give the definitions of the Passarino--Veltman three point 
functions with the convention of \cite{Denner}:
\be
{\mathcal D}^{0} = q^{2} - m_{0}^{2}
\quad \mbox{and}\quad
{\mathcal D}^{j} = ( q + p_{j} )^{2} - m_{j}^{2}
\ee
are the general denominators for
\be & &
  C_{0}(p_{1}^{2},(p_{1}-p_{2})^{2},p_{2}^{2}
       ,m_{0}^{2},m_{1}^{2},m_{2}^{2})
  :=
  \frac{1}{i \pi^{2}} \int d^{D} q
\frac{1}{{\mathcal D}^{0} {\mathcal D}^{1} {\mathcal D}^{2}}
\\  & &
  C_{\mu}(p_{1}^{2},(p_{1}-p_{2})^{2},p_{2}^{2}
         ,m_{0}^{2},m_{1}^{2},m_{2}^{2})
  :=
  \frac{1}{i \pi^{2}} \int d^{D} q
\frac{q_{\mu}}{{\mathcal D}^{0} {\mathcal D}^{1} {\mathcal D}^{2}}
\\  & & \quad
  = \enspace\,
    p_{1\mu} C_{1}(p_{1}^{2},(p_{1}-p_{2})^{2},p_{2}^{2}
                  ,m_{0}^{2},m_{1}^{2},m_{2}^{2})
\nn \\  & & \qquad
  + p_{2\mu} C_{2}(p_{1}^{2},(p_{1}-p_{2})^{2},p_{2}^{2}
                  ,m_{0}^{2},m_{1}^{2},m_{2}^{2})
\nn \\ & &
  C_{\mu\nu}(p_{1}^{2},(p_{1}-p_{2})^{2},p_{2}^{2}
            ,m_{0}^{2},m_{1}^{2},m_{2}^{2})
  :=
  \frac{1}{i \pi^{2}} \int d^{D} q
\frac{q_{\mu} q_{\nu}}
 {{\mathcal D}^{0} {\mathcal D}^{1} {\mathcal D}^{2}}
\\  & & \quad
  = \enspace\,
    g_{\mu\nu}
C_{00}(p_{1}^{2},(p_{1}-p_{2})^{2},p_{2}^{2}
      ,m_{0}^{2},m_{1}^{2},m_{2}^{2})
\nn \\ & & \qquad
  + p_{1\mu} p_{1\nu}
C_{11}(p_{1}^{2},(p_{1}-p_{2})^{2},p_{2}^{2}
      ,m_{0}^{2},m_{1}^{2},m_{2}^{2})
\nn \\ & & \qquad
  + ( p_{1\mu} p_{2\nu} + p_{2\mu} p_{1\nu} )
C_{12}(p_{1}^{2},(p_{1}-p_{2})^{2},p_{2}^{2}
      ,m_{0}^{2},m_{1}^{2},m_{2}^{2})
\nn \\ & & \qquad
  + p_{2\mu} p_{2\nu}
C_{22}(p_{1}^{2},(p_{1}-p_{2})^{2},p_{2}^{2}
      ,m_{0}^{2},m_{1}^{2},m_{2}^{2})
\; .
\nn \ee
We further use the definition
\be
C_{x}^{m,jk} = 
C_{x} ( m_{t}^{2}, s , m_{t}^{2}
      , m_{\tilde{t}_{m}}^{2}
      , m_{\tilde{\chi}^{0}_{j}}^{2}
      , m_{\tilde{\chi}^{0}_{k}}^{2} ) 
\quad
x \in \{ 0, 1, 2, 11, 22 \}
\ee
and it follows
\be
C_{0}^{m,jk} = C_{0}^{m,kj}
\qquad
C_{1}^{m,jk} = C_{2}^{m,kj}
\qquad
C_{11}^{m,jk} = C_{22}^{m,kj}
\; .
\ee

\section*{Figure Captions}

\paragraph{Figure~1:} 
Feynman diagrams contributing to $\dgs$ and $\dzs$: 
\\ \textbf{(a)} with two fermions and one scalar in the loop
\\ \textbf{(b)} with two scalars and one fermion in the loop.

\paragraph{Figure~2:} 
Dependence of the masses (in GeV) on \mbox{$\cos\varphi_{\mu}$} for 
\mbox{$M = 230$~GeV}. 
\\ \textbf{(a)} charginos: 
$m_{\tilde{\chi}^{+}_{1}}$ (full line), 
$m_{\tilde{\chi}^{+}_{2}}$ (dashed line).
\\ \textbf{(b)} neutralinos: 
$m_{\tilde{\chi}^{0}_{1}}$ (full line), 
$m_{\tilde{\chi}^{0}_{2}}$ (dashed line),
\\ \mbox{\hspace{6.3mm}}
$m_{\tilde{\chi}^{0}_{3}}$ (dotted line), 
$m_{\tilde{\chi}^{0}_{4}}$ (dashed--dotted line).

\paragraph{Figure~3:} 
Dependence of the chargino contribution to $\dgs$ and $\dzs$ on 
$|\mu|$~(GeV) for the reference parameter set. 
\\ \textbf{(a)} 
\mbox{$\mIm d^{\gamma}_{\tilde{\chi}^{+}}$} for 
\mbox{$M = 230$~GeV} (full line), 
\mbox{$M = 360$~GeV} (dotted line), 
\\ \mbox{\hspace{6.3mm}}
\mbox{$\mIm d^{Z}_{\tilde{\chi}^{+}}$} for 
\mbox{$M = 230$~GeV} (dashed line), 
\mbox{$M = 360$~GeV} (dashed--dotted line). 
\\ \textbf{(b)} 
\mbox{$\eRe d^{\gamma}_{\tilde{\chi}^{+}}$} for 
\mbox{$M = 230$~GeV} (full line), 
\mbox{$M = 360$~GeV} (dotted line), 
\\ \mbox{\hspace{6.3mm}}
\mbox{$\eRe d^{Z}_{\tilde{\chi}^{+}}$} for 
\mbox{$M = 230$~GeV} (dashed line), 
\mbox{$M = 360$~GeV} (dashed--dotted line). 

\paragraph{Figure~4:} 
Dependence of the neutralino contribution to $\dgs$ and $\dzs$ on 
$|\mu|$~(GeV) for the reference parameter set. 
\\ \textbf{(a)} 
\mbox{$\mIm d^{\gamma}_{\tilde{\chi}^{0}}$} for 
\mbox{$M = 230$~GeV} (full line), 
\mbox{$M = 360$~GeV} (dotted line), 
\\ \mbox{\hspace{6.3mm}}
\mbox{$\mIm d^{Z}_{\tilde{\chi}^{0}}$} for 
\mbox{$M = 230$~GeV} (dashed line), 
\mbox{$M = 360$~GeV} (dashed--dotted line). 
\\ \textbf{(b)} 
\mbox{$\eRe d^{\gamma}_{\tilde{\chi}^{0}}$} for 
\mbox{$M = 230$~GeV} (full line), 
\mbox{$M = 360$~GeV} (dotted line), 
\\ \mbox{\hspace{6.3mm}}
\mbox{$\eRe d^{Z}_{\tilde{\chi}^{0}}$} for 
\mbox{$M = 230$~GeV} (dashed line), 
\mbox{$M = 360$~GeV} (dashed--dotted line). 

\paragraph{Figure~5:} 
Dependence of the chargino/neutralino contributions to 
$\dgs$ and $\dzs$ on $\varphi_{\mu}$ for the reference parameter set. 
\\ \textbf{(a)} 
\mbox{$\mIm d^{\gamma}_{\tilde{\chi}^{+}}$} for 
\mbox{$M = 230$~GeV} (full line), 
\mbox{$M = 360$~GeV} (dotted line), 
\\ \mbox{\hspace{6.3mm}}
\mbox{$\mIm d^{Z}_{\tilde{\chi}^{+}}$} for 
\mbox{$M = 230$~GeV} (dashed line), 
\mbox{$M = 360$~GeV} (dashed--dotted line). 
\\ \textbf{(b)} 
\mbox{$\eRe d^{\gamma}_{\tilde{\chi}^{+}}$} for 
\mbox{$M = 230$~GeV} (full line), 
\mbox{$M = 360$~GeV} (dotted line), 
\\ \mbox{\hspace{6.3mm}}
\mbox{$\eRe d^{Z}_{\tilde{\chi}^{+}}$} for 
\mbox{$M = 230$~GeV} (dashed line), 
\mbox{$M = 360$~GeV} (dashed--dotted line). 
\\ \textbf{(c)} 
\mbox{$\mIm d^{\gamma}_{\tilde{\chi}^{0}}$} for 
\mbox{$M = 230$~GeV} (full line), 
\mbox{$M = 360$~GeV} (dotted line), 
\\ \mbox{\hspace{6.3mm}}
\mbox{$\mIm d^{Z}_{\tilde{\chi}^{0}}$} for 
\mbox{$M = 230$~GeV} (dashed line), 
\mbox{$M = 360$~GeV} (dashed--dotted line). 
\\ \textbf{(d)}
\mbox{$\eRe d^{\gamma}_{\tilde{\chi}^{0}}$} for 
\mbox{$M = 230$~GeV} (full line), 
\mbox{$M = 360$~GeV} (dotted line), 
\\ \mbox{\hspace{6.3mm}}
\mbox{$\eRe d^{Z}_{\tilde{\chi}^{0}}$} for 
\mbox{$M = 230$~GeV} (dashed line), 
\mbox{$M = 360$~GeV} (dashed--dotted line). 

\paragraph{Figure~6:} 
$\dgs$ and $\dzs$ for the reference parameter set with 
\mbox{$M = 230$~GeV}.
\\ \textbf{(a)} 
\mbox{$\mIm \dgs$} (full line), 
\mbox{$\mIm \dzs$} (dashed line)
\\ \textbf{(b)} 
\mbox{$\eRe \dgs$} (full line), 
\mbox{$\eRe \dzs$} (dashed line).

\newpage 

\input figures.tex

\end{document}

%% file: figures.tex
\begin{center}
\setlength{\unitlength}{1mm}
\begin{picture}(140,70)
\put(  0, 5){\mbox{\epsfig{file=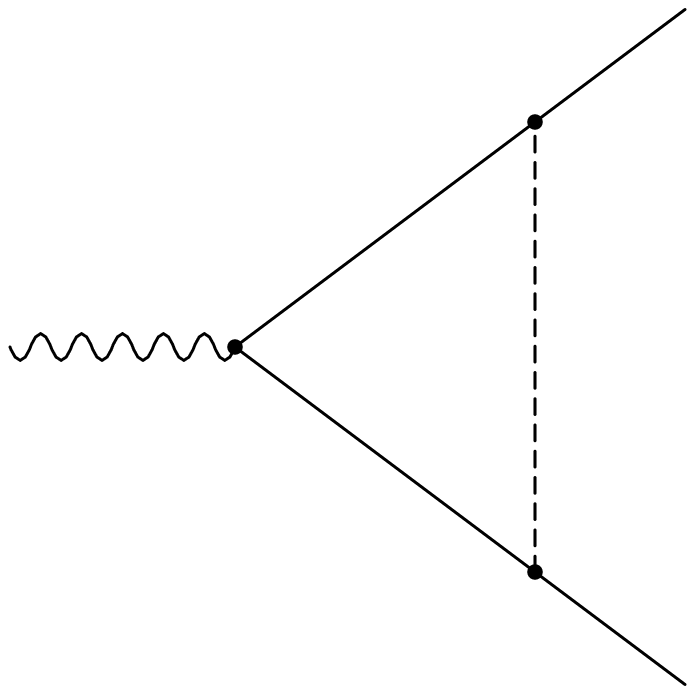,width=6cm}}}
\put( 10,37){\makebox(0,0)[b ]{$A^{\mu}, Z^{\mu}$}}
\put( 60,62){\makebox(0,0)[tl]{$\bar{t}$}}
\put( 60, 6){\makebox(0,0)[bl]{$t$}}
\put( 35,46){\makebox(0,0)[br]{$\overline{\tilde{\chi}}^{0}_{k}, 
                                \overline{\tilde{\chi}}^{+}_{k}$}}
\put( 35,24){\makebox(0,0)[tr]{$\tilde{\chi}^{0}_{j}, 
                                \tilde{\chi}^{+}_{j}$}}
\put( 47,34){\makebox(0,0)[ l]{$\begin{array}{l}
                                  \tilde{t}_{m} \, , \\ 
                                  \tilde{b}_{m}
                                \end{array}$}}
\put( 30, 0){\makebox(0,0)[b ]{{\large\bf Fig.~1a}}}
\put(  70, 5){\mbox{\epsfig{file=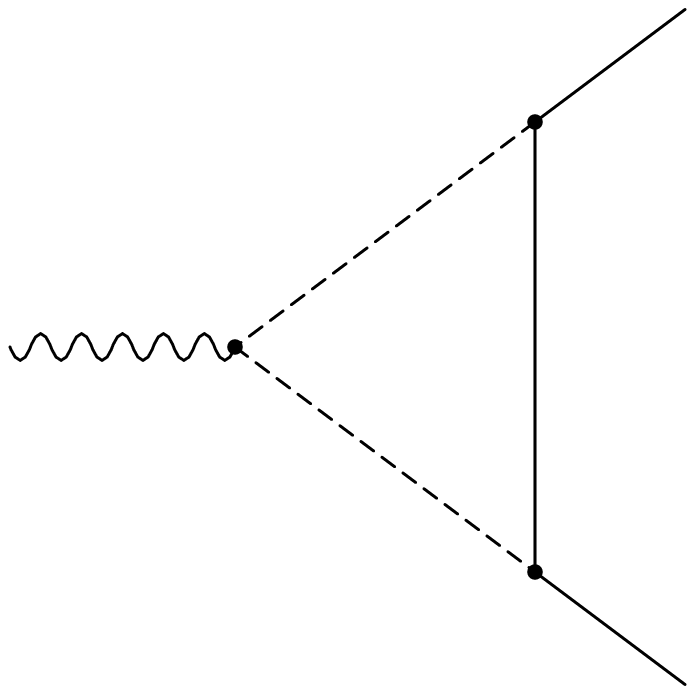,width=6cm}}}
\put( 80,37){\makebox(0,0)[b ]{$A^{\mu}, Z^{\mu}$}}
\put(130,62){\makebox(0,0)[tl]{$\bar{t}$}}
\put(130, 6){\makebox(0,0)[bl]{$t$}}
\put(105,46){\makebox(0,0)[br]{$\tilde{t}_{m}^{*}, \tilde{b}_{m}^{*}$}}
\put(105,24){\makebox(0,0)[tr]{$\tilde{t}_{n}, \tilde{b}_{n}$}}
\put(117,34){\makebox(0,0)[ l]{$\begin{array}{l}
                                  \tilde{g} \, , \\ 
                                  \tilde{\chi}^{0}_{k} \, , \\ 
                                  \tilde{\chi}^{+}_{k}
                                \end{array}$}}
\put(100, 0){\makebox(0,0)[b ]{{\large\bf Fig.~1b}}}
\end{picture}\\
\setlength{\unitlength}{1pt}
\end{center}

\begin{center}
\setlength{\unitlength}{1mm}
\begin{picture}(140,120)(10,0)
\put( 7,15){\mbox{\epsfig{file=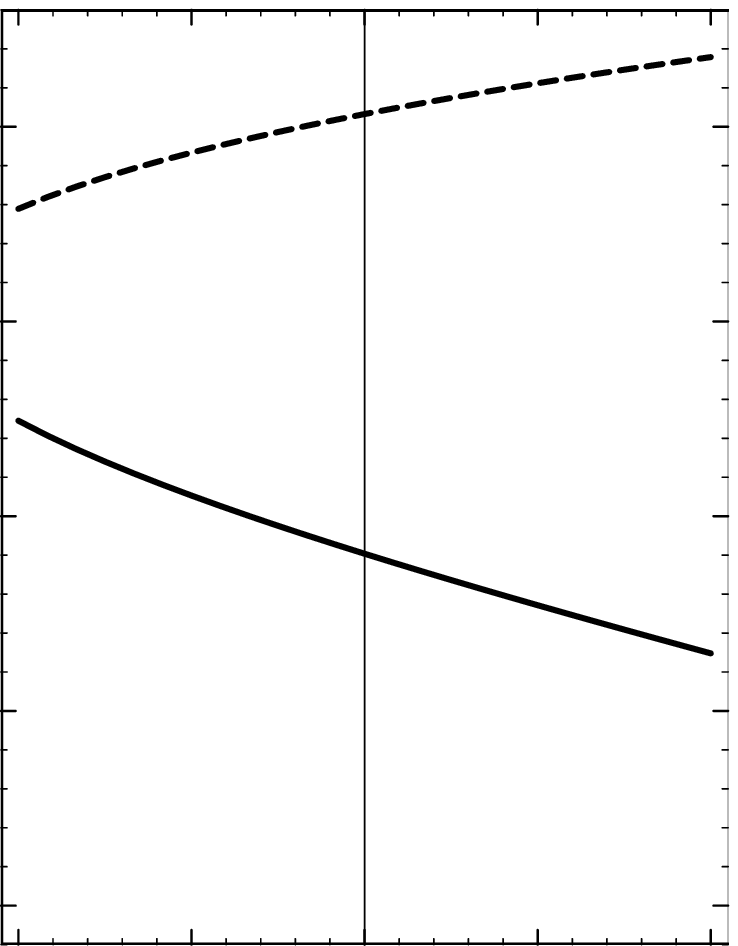,width=63mm}}}
\put( 53.5,13){\makebox(0,0)[t ]{$\cos \varphi_{\mu}$}}
\put( 35,50){\makebox(0,0)[tr]{$m_{\tilde{\chi}^{+}_{1}}$}}
\put( 35,82){\makebox(0,0)[tr]{$m_{\tilde{\chi}^{+}_{2}}$}}
\put(  6,19   ){\makebox(0,0)[ r]{\small 100}}
\put(  6,35.75){\makebox(0,0)[ r]{\small 150}}
\put(  6,52.50){\makebox(0,0)[ r]{\small 200}}
\put(  6,69.25){\makebox(0,0)[ r]{\small 250}}
\put(  6,86   ){\makebox(0,0)[ r]{\small 300}}
\put(  6,92.7 ){\makebox(0,0)[ r]{\small GeV}}
\put(  8.5,14){\makebox(0,0)[t]{\small -1}}
\put( 38.5,14){\makebox(0,0)[t]{\small  0}}
\put( 68.5,14){\makebox(0,0)[t]{\small  1}}
\put( 38, 2){\makebox(0,0)[b ]{\large\bf Fig.~2a}}
\put(70,15){\mbox{\epsfig{file=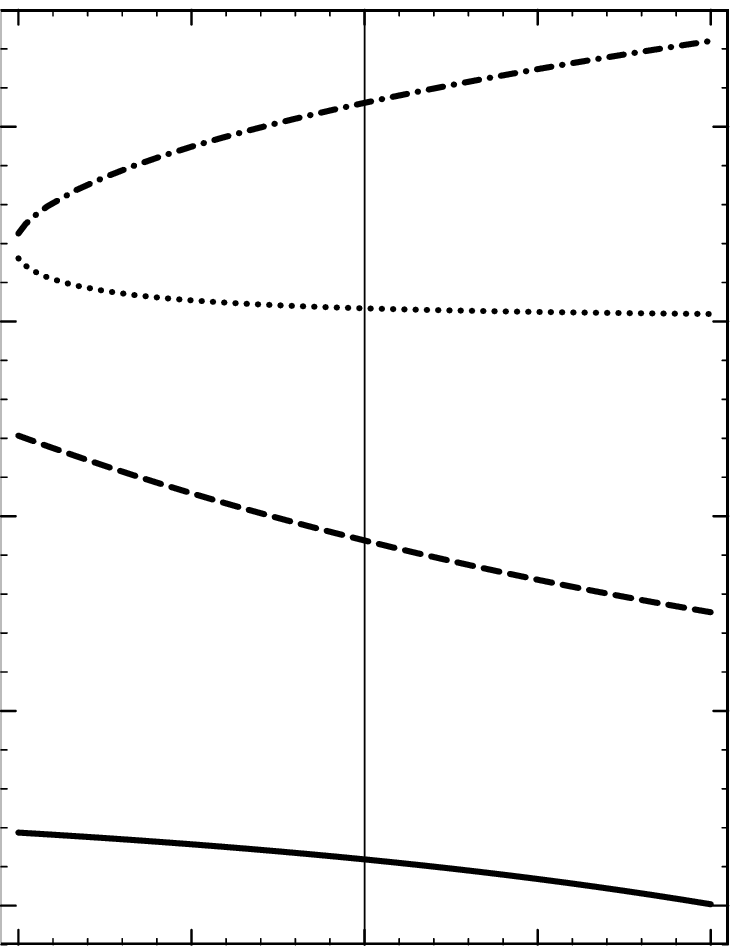,width=63mm}}}
\put(116.5,13){\makebox(0,0)[t ]{$\cos \varphi_{\mu}$}}
\put(110,26){\makebox(0,0)[tl]{$m_{\tilde{\chi}^{0}_{1}}$}}
\put(110,46){\makebox(0,0)[tl]{$m_{\tilde{\chi}^{0}_{2}}$}}
\put(110,68){\makebox(0,0)[tl]{$m_{\tilde{\chi}^{0}_{3}}$}}
\put(110,88){\makebox(0,0)[tl]{$m_{\tilde{\chi}^{0}_{4}}$}}
\put(101, 2){\makebox(0,0)[b ]{\large\bf Fig.~2b}}
\put(134,19   ){\makebox(0,0)[ l]{\small 100}}
\put(134,35.75){\makebox(0,0)[ l]{\small 150}}
\put(134,52.50){\makebox(0,0)[ l]{\small 200}}
\put(134,69.25){\makebox(0,0)[ l]{\small 250}}
\put(134,86   ){\makebox(0,0)[ l]{\small 300}}
\put(134,92.7 ){\makebox(0,0)[ l]{\small GeV}}
\put( 71.5,14){\makebox(0,0)[t]{\small -1}}
\put(101.5,14){\makebox(0,0)[t]{\small  0}}
\put(131.5,14){\makebox(0,0)[t]{\small  1}}
\end{picture}\\
\setlength{\unitlength}{1pt}
\end{center}

\begin{center}
\setlength{\unitlength}{1mm}
\begin{picture}(140,90)(10,0)
\put( 7,15){\mbox{\epsfig{file=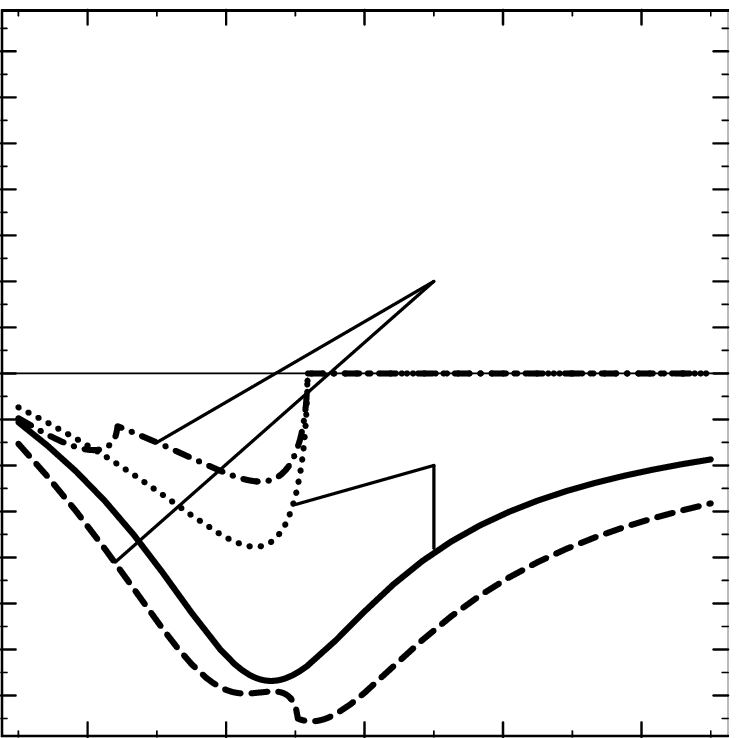,width=63mm}}}
\put( 45,54.5){\makebox(0,0)[ l]{$\mIm d^{Z}_{\tilde{\chi}^{+}}$}}
\put( 45,40.5){\makebox(0,0)[ l]{$\mIm d^{\gamma}_{\tilde{\chi}^{+}}$}}
\put(  6,70.1){\makebox(0,0)[ r]{\small  0.0006}}
\put(  6,62.3){\makebox(0,0)[ r]{\small  0.0004}}
\put(  6,54.5){\makebox(0,0)[ r]{\small  0.0002}}
\put(  6,46.7){\makebox(0,0)[ r]{\small  0}}
\put(  6,38.9){\makebox(0,0)[ r]{\small -0.0002}}
\put(  6,31.1){\makebox(0,0)[ r]{\small -0.0004}}
\put(  6,23.3){\makebox(0,0)[ r]{\small -0.0006}}
\put( 14,14){\makebox(0,0)[t]{\small 100}}
\put( 26,14){\makebox(0,0)[t]{\small 200}}
\put( 38,14){\makebox(0,0)[t]{\small 300}}
\put( 50,14){\makebox(0,0)[t]{\small 400}}
\put( 62,14){\makebox(0,0)[t]{\small 500}}
\put( 38, 2){\makebox(0,0)[b ]{\large\bf Fig.~3a}}
\put(70,15){\mbox{\epsfig{file=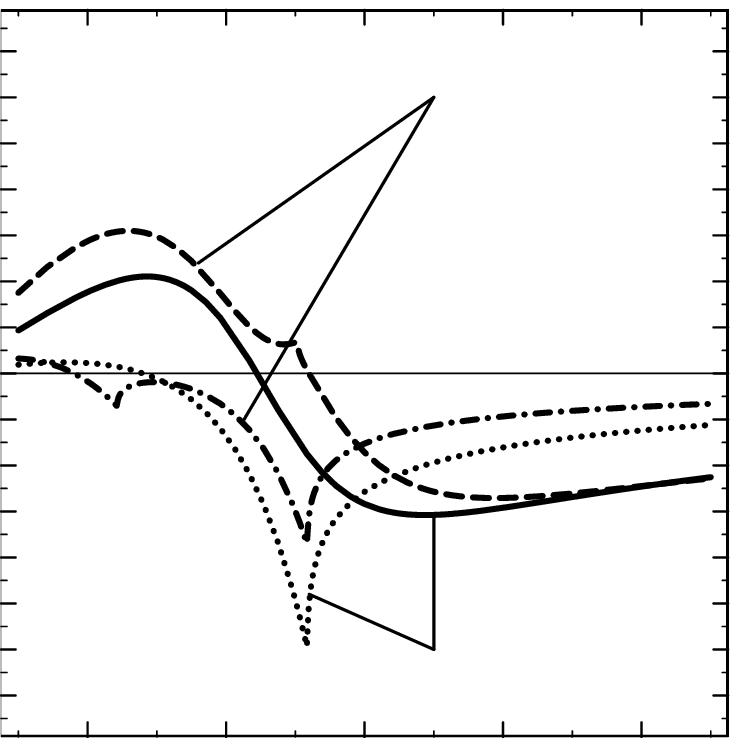,width=63mm}}}
\put( 70, 3){\makebox(0,0)[b ]{$|\mu|/$GeV}}
\put(108,70.1){\makebox(0,0)[ l]{$\eRe d^{Z}_{\tilde{\chi}^{+}}$}}
\put(108,23.3){\makebox(0,0)[ l]{$\eRe d^{\gamma}_{\tilde{\chi}^{+}}$}}
\put(134,70.1){\makebox(0,0)[ l]{\small  0.0006}}
\put(134,62.3){\makebox(0,0)[ l]{\small  0.0004}}
\put(134,54.5){\makebox(0,0)[ l]{\small  0.0002}}
\put(134,46.7){\makebox(0,0)[ l]{\small  0}}
\put(134,38.9){\makebox(0,0)[ l]{\small -0.0002}}
\put(134,31.1){\makebox(0,0)[ l]{\small -0.0004}}
\put(134,23.3){\makebox(0,0)[ l]{\small -0.0006}}
\put( 77,14){\makebox(0,0)[t]{\small 100}}
\put( 89,14){\makebox(0,0)[t]{\small 200}}
\put(101,14){\makebox(0,0)[t]{\small 300}}
\put(113,14){\makebox(0,0)[t]{\small 400}}
\put(125,14){\makebox(0,0)[t]{\small 500}}
\put(101, 2){\makebox(0,0)[b ]{\large\bf Fig.~3b}}
\end{picture}\\
\setlength{\unitlength}{1pt}
\end{center}

\begin{center}
\setlength{\unitlength}{1mm}
\begin{picture}(140,90)(10,0)
\put( 7,15){\mbox{\epsfig{file=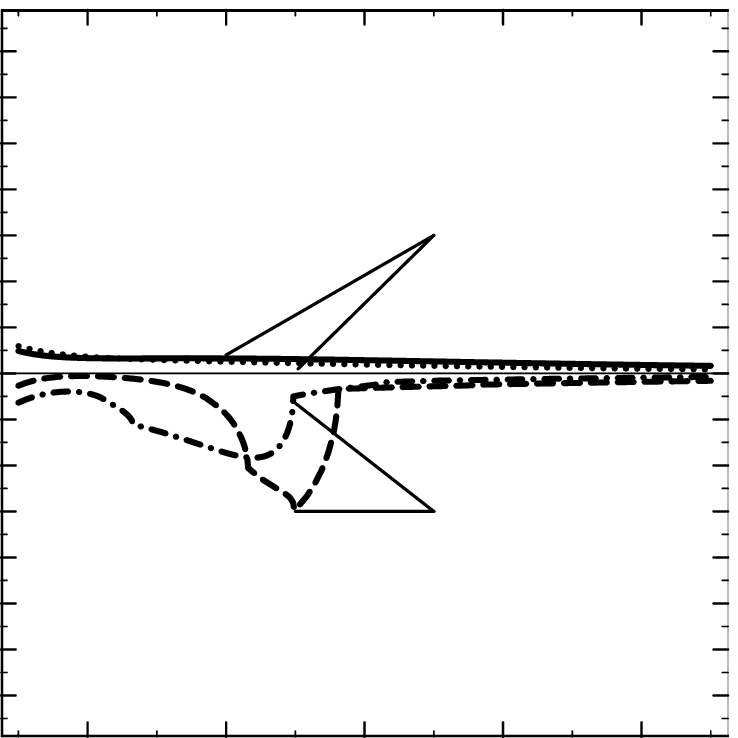,width=63mm}}}
\put( 45,33){\makebox(0,0)[ l]{$\mIm d^{Z}_{\tilde{\chi}^{0}}$}}
\put( 45,58.4){\makebox(0,0)[ l]{$\mIm d^{\gamma}_{\tilde{\chi}^{0}}$}}
\put(  6,70.1){\makebox(0,0)[ r]{\small  0.0006}}
\put(  6,62.3){\makebox(0,0)[ r]{\small  0.0004}}
\put(  6,54.5){\makebox(0,0)[ r]{\small  0.0002}}
\put(  6,46.7){\makebox(0,0)[ r]{\small  0}}
\put(  6,38.9){\makebox(0,0)[ r]{\small -0.0002}}
\put(  6,31.1){\makebox(0,0)[ r]{\small -0.0004}}
\put(  6,23.3){\makebox(0,0)[ r]{\small -0.0006}}
\put( 14,14){\makebox(0,0)[t]{\small 100}}
\put( 26,14){\makebox(0,0)[t]{\small 200}}
\put( 38,14){\makebox(0,0)[t]{\small 300}}
\put( 50,14){\makebox(0,0)[t]{\small 400}}
\put( 62,14){\makebox(0,0)[t]{\small 500}}
\put( 38, 2){\makebox(0,0)[b ]{\large\bf Fig.~4a}}
\put( 70, 3){\makebox(0,0)[b ]{$|\mu|/$GeV}}
\put(70,15){\mbox{\epsfig{file=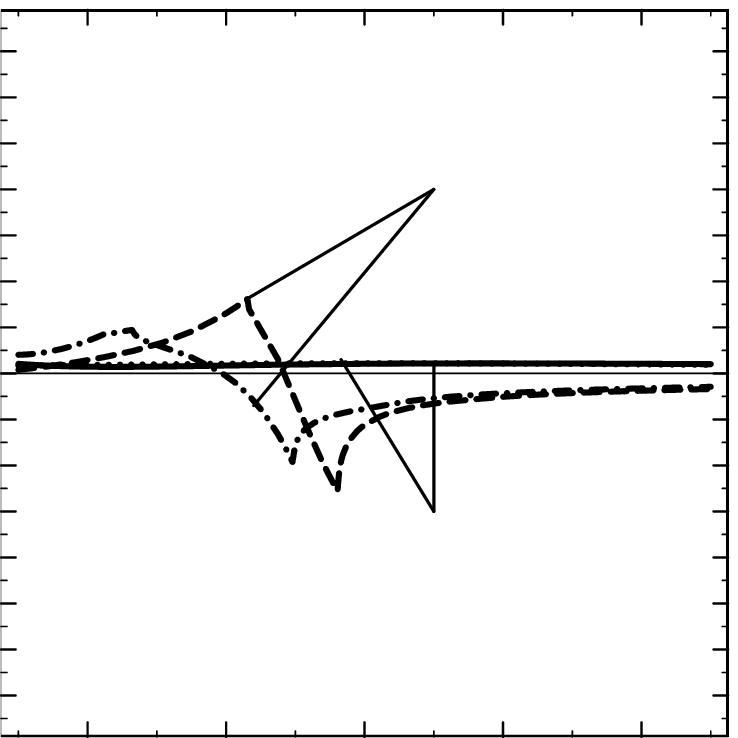,width=63mm}}}
\put(108,33){\makebox(0,0)[ l]{$\eRe d^{\gamma}_{\tilde{\chi}^{0}}$}}
\put(108,62.3){\makebox(0,0)[ l]{$\eRe d^{Z}_{\tilde{\chi}^{0}}$}}
\put(134,70.1){\makebox(0,0)[ l]{\small  0.0006}}
\put(134,62.3){\makebox(0,0)[ l]{\small  0.0004}}
\put(134,54.5){\makebox(0,0)[ l]{\small  0.0002}}
\put(134,46.7){\makebox(0,0)[ l]{\small  0}}
\put(134,38.9){\makebox(0,0)[ l]{\small -0.0002}}
\put(134,31.1){\makebox(0,0)[ l]{\small -0.0004}}
\put(134,23.3){\makebox(0,0)[ l]{\small -0.0006}}
\put( 78,14){\makebox(0,0)[t]{\small 100}}
\put( 89,14){\makebox(0,0)[t]{\small 200}}
\put(102,14){\makebox(0,0)[t]{\small 300}}
\put(113,14){\makebox(0,0)[t]{\small 400}}
\put(125,14){\makebox(0,0)[t]{\small 500}}
\put(101, 2){\makebox(0,0)[b ]{\large\bf Fig.~4b}}
\end{picture}\\
\setlength{\unitlength}{1pt}
\end{center}

\begin{center}
\setlength{\unitlength}{1mm}
\begin{picture}(140,90)(10,0)
\put( 7,15){\mbox{\epsfig{file=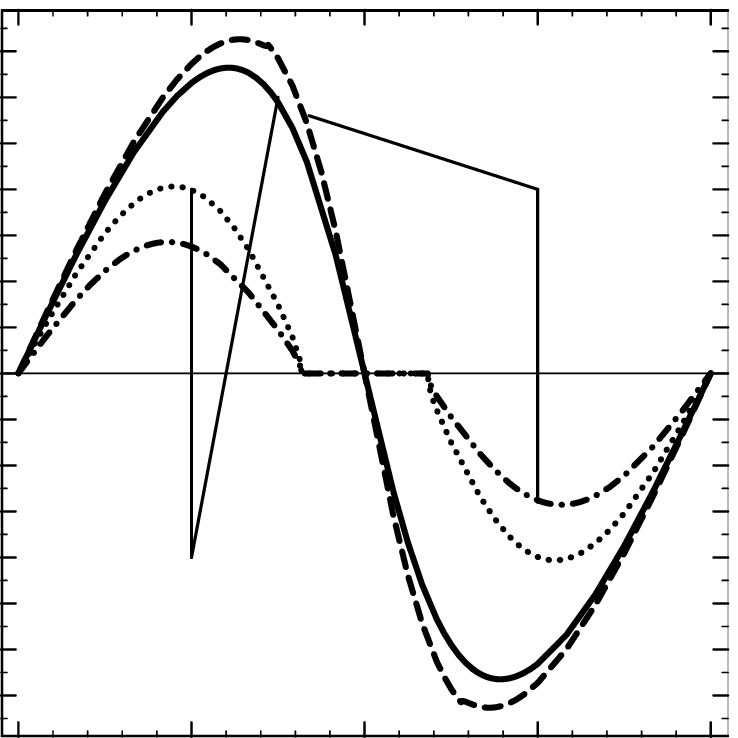,width=63mm}}}
\put( 54,62.3){\makebox(0,0)[ l]{$\mIm d^{Z}_{\tilde{\chi}^{+}}$}}
\put( 24,31){\makebox(0,0)[t ]{$\mIm d^{\gamma}_{\tilde{\chi}^{+}}$}}
\put(  6,70.1){\makebox(0,0)[ r]{\small  0.0006}}
\put(  6,62.3){\makebox(0,0)[ r]{\small  0.0004}}
\put(  6,54.5){\makebox(0,0)[ r]{\small  0.0002}}
\put(  6,46.7){\makebox(0,0)[ r]{\small  0}}
\put(  6,38.9){\makebox(0,0)[ r]{\small -0.0002}}
\put(  6,31.1){\makebox(0,0)[ r]{\small -0.0004}}
\put(  6,23.3){\makebox(0,0)[ r]{\small -0.0006}}
\put(  9  ,10){\makebox(0,0)[bl]{\small 0}}
\put( 38.5,10){\makebox(0,0)[b ]{\small $\pi$}}
\put( 68  ,10){\makebox(0,0)[br]{\small $2\pi$}}
\put( 53,  11){\makebox(0,0)[ ]{$\varphi_{\mu} \to$}}
\put( 38, 2){\makebox(0,0)[b ]{\large\bf Fig.~5a}}
\put(70,15){\mbox{\epsfig{file=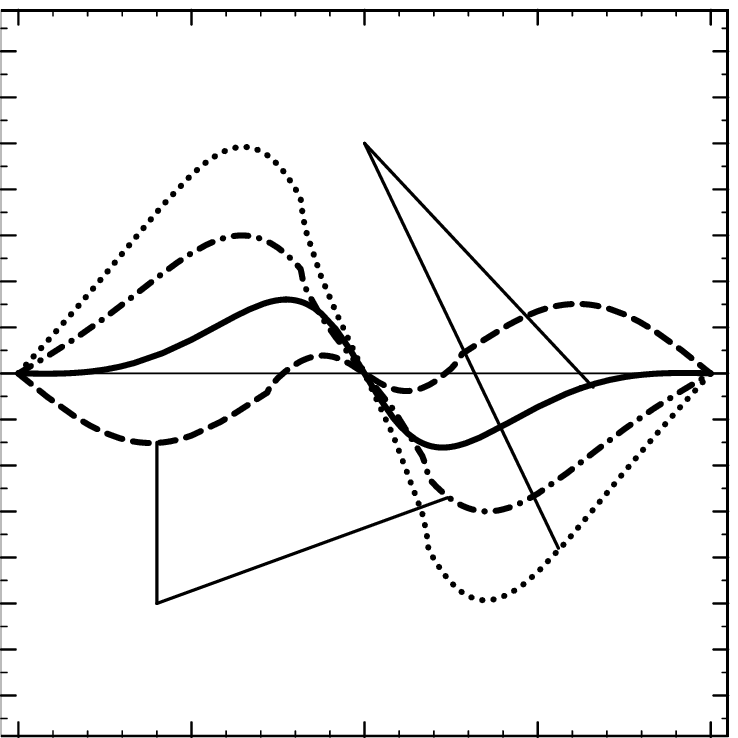,width=63mm}}}
\put( 86  ,27.1){\makebox(0,0)[t ]{$\eRe d^{Z}_{\tilde{\chi}^{+}}$}}
\put(101.5,66.5){\makebox(0,0)[b ]{$\eRe d^{\gamma}_{\tilde{\chi}^{+}}$}}
\put(134,70.1){\makebox(0,0)[ l]{\small  0.0006}}
\put(134,62.3){\makebox(0,0)[ l]{\small  0.0004}}
\put(134,54.5){\makebox(0,0)[ l]{\small  0.0002}}
\put(134,46.7){\makebox(0,0)[ l]{\small  0}}
\put(134,38.9){\makebox(0,0)[ l]{\small -0.0002}}
\put(134,31.1){\makebox(0,0)[ l]{\small -0.0004}}
\put(134,23.3){\makebox(0,0)[ l]{\small -0.0006}}
\put( 72  ,10){\makebox(0,0)[bl]{\small 0}}
\put(101.5,10){\makebox(0,0)[b ]{\small $\pi$}}
\put(131  ,10){\makebox(0,0)[br]{\small $2\pi$}}
\put(116,  11){\makebox(0,0)[ ]{$\varphi_{\mu} \to$}}
\put(101, 2){\makebox(0,0)[b ]{\large\bf Fig.~5b}}
\end{picture}\\
\setlength{\unitlength}{1pt}
\end{center}

\begin{center}
\setlength{\unitlength}{1mm}
\begin{picture}(140,90)(10,0)
\put( 7,15){\mbox{\epsfig{file=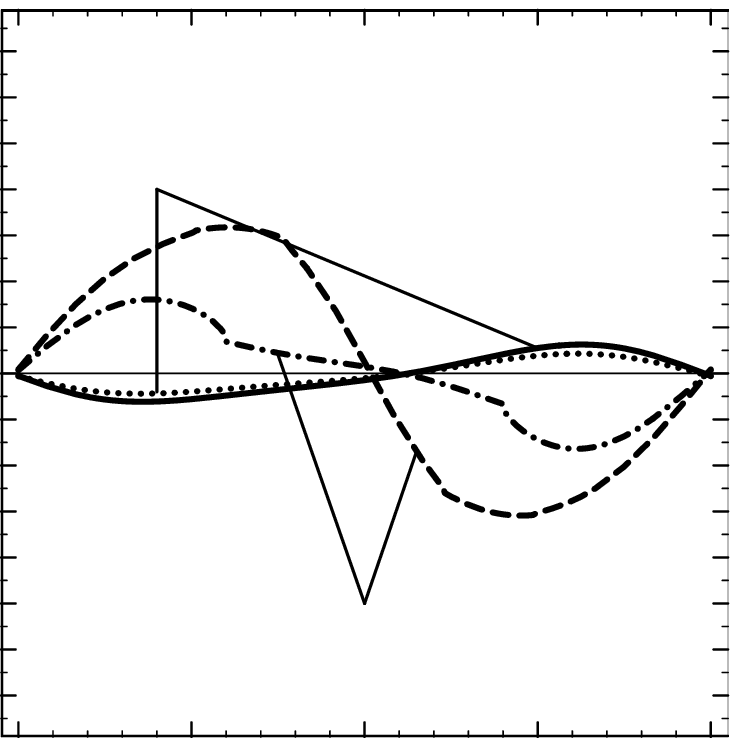,width=63mm}}}
\put( 39,26.7){\makebox(0,0)[t ]{$\mIm d^{Z}_{\tilde{\chi}^{0}}$}}
\put( 21,63.2){\makebox(0,0)[b ]{$\mIm d^{\gamma}_{\tilde{\chi}^{0}}$}}
\put(  6,70.1){\makebox(0,0)[ r]{\small  0.0006}}
\put(  6,62.3){\makebox(0,0)[ r]{\small  0.0004}}
\put(  6,54.5){\makebox(0,0)[ r]{\small  0.0002}}
\put(  6,46.7){\makebox(0,0)[ r]{\small  0}}
\put(  6,38.9){\makebox(0,0)[ r]{\small -0.0002}}
\put(  6,31.1){\makebox(0,0)[ r]{\small -0.0004}}
\put(  6,23.3){\makebox(0,0)[ r]{\small -0.0006}}
\put(  9  ,10){\makebox(0,0)[bl]{\small 0}}
\put( 38.5,10){\makebox(0,0)[b ]{\small $\pi$}}
\put( 68  ,10){\makebox(0,0)[br]{\small $2\pi$}}
\put( 53,  11){\makebox(0,0)[ ]{$\varphi_{\mu} \to$}}
\put( 38, 2){\makebox(0,0)[b ]{\large\bf Fig.~5c}}
\put(70,15){\mbox{\epsfig{file=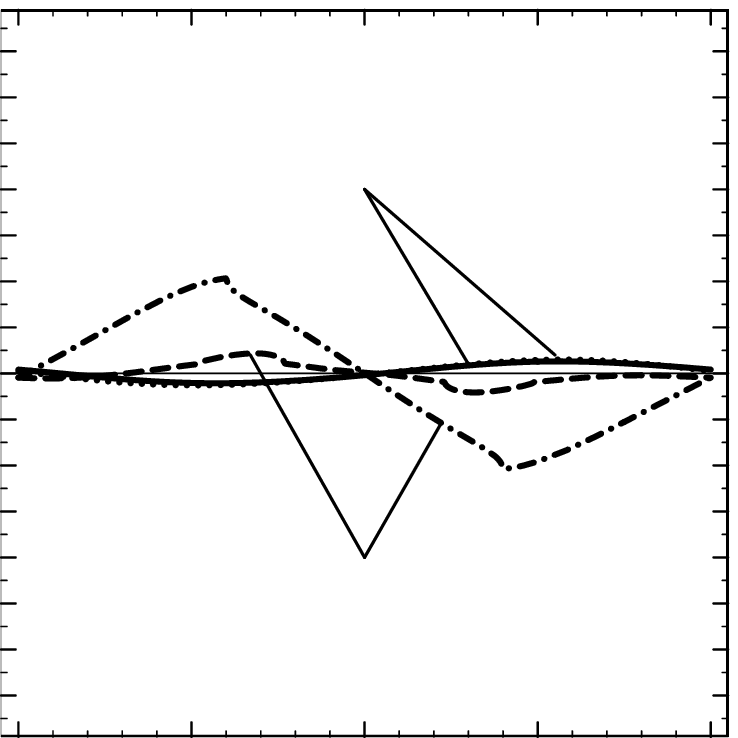,width=63mm}}}
\put(101.5,31.1){\makebox(0,0)[t ]{$\eRe d^{Z}_{\tilde{\chi}^{0}}$}}
\put(101.5,62.7){\makebox(0,0)[b ]{$\eRe d^{\gamma}_{\tilde{\chi}^{0}}$}}
\put(134,70.1){\makebox(0,0)[ l]{\small  0.0006}}
\put(134,62.3){\makebox(0,0)[ l]{\small  0.0004}}
\put(134,54.5){\makebox(0,0)[ l]{\small  0.0002}}
\put(134,46.7){\makebox(0,0)[ l]{\small  0}}
\put(134,38.9){\makebox(0,0)[ l]{\small -0.0002}}
\put(134,31.1){\makebox(0,0)[ l]{\small -0.0004}}
\put(134,23.3){\makebox(0,0)[ l]{\small -0.0006}}
\put( 72  ,10){\makebox(0,0)[bl]{\small 0}}
\put(101.5,10){\makebox(0,0)[b ]{\small $\pi$}}
\put(131  ,10){\makebox(0,0)[br]{\small $2\pi$}}
\put(116,  11){\makebox(0,0)[ ]{$\varphi_{\mu} \to$}}
\put(101, 2){\makebox(0,0)[b ]{\large\bf Fig.~5d}}
\end{picture}\\
\setlength{\unitlength}{1pt}
\end{center}

\begin{center}
\setlength{\unitlength}{1mm}
\begin{picture}(140,120)(10,0)
\put( 7,15){\mbox{\epsfig{file=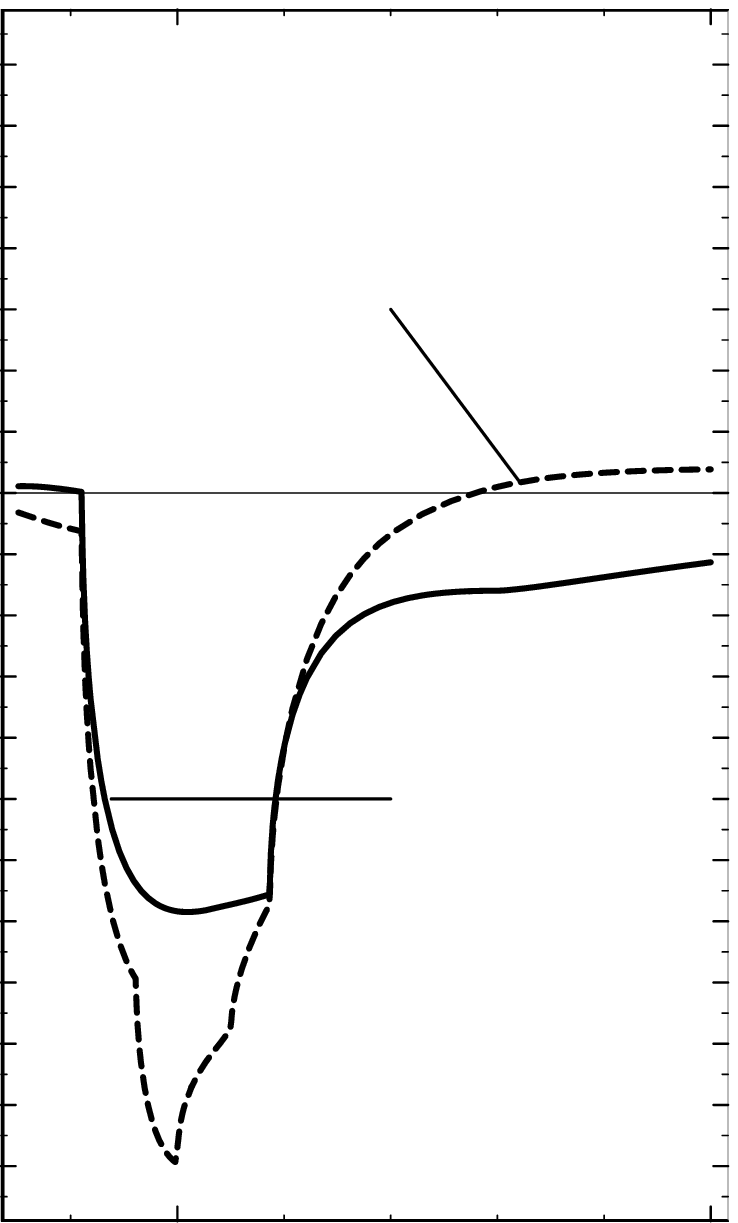,width=63mm}}}
\put( 41,94.5){\makebox(0,0)[b ]{$\mIm \dzs$}}
\put( 42,52.1){\makebox(0,0)[ l]{$\mIm \dgs$}}
\put(  6,110.4){\makebox(0,0)[ r]{\small  0.0006}}
\put(  6, 99.8){\makebox(0,0)[ r]{\small  0.0004}}
\put(  6, 89.2){\makebox(0,0)[ r]{\small  0.0002}}
\put(  6, 78.6){\makebox(0,0)[ r]{\small  0}}
\put(  6, 68.0){\makebox(0,0)[ r]{\small -0.0002}}
\put(  6, 57.4){\makebox(0,0)[ r]{\small -0.0004}}
\put(  6, 46.8){\makebox(0,0)[ r]{\small -0.0006}}
\put(  6, 36.2){\makebox(0,0)[ r]{\small -0.0008}}
\put(  6, 25.6){\makebox(0,0)[ r]{\small -0.0010}}
\put( 13  ,14){\makebox(0,0)[t]{\small 400}}
\put( 22.2,14){\makebox(0,0)[t]{\small 500}}
\put( 31.4,14){\makebox(0,0)[t]{\small 600}}
\put( 40.6,14){\makebox(0,0)[t]{\small 700}}
\put( 49.8,14){\makebox(0,0)[t]{\small 800}}
\put( 59  ,14){\makebox(0,0)[t]{\small 900}}
\put( 70, 3){\makebox(0,0)[b ]{$\sqrt{s}/$GeV}}
\put( 38, 2){\makebox(0,0)[b ]{\large\bf Fig.~6a}}
\put(70,15){\mbox{\epsfig{file=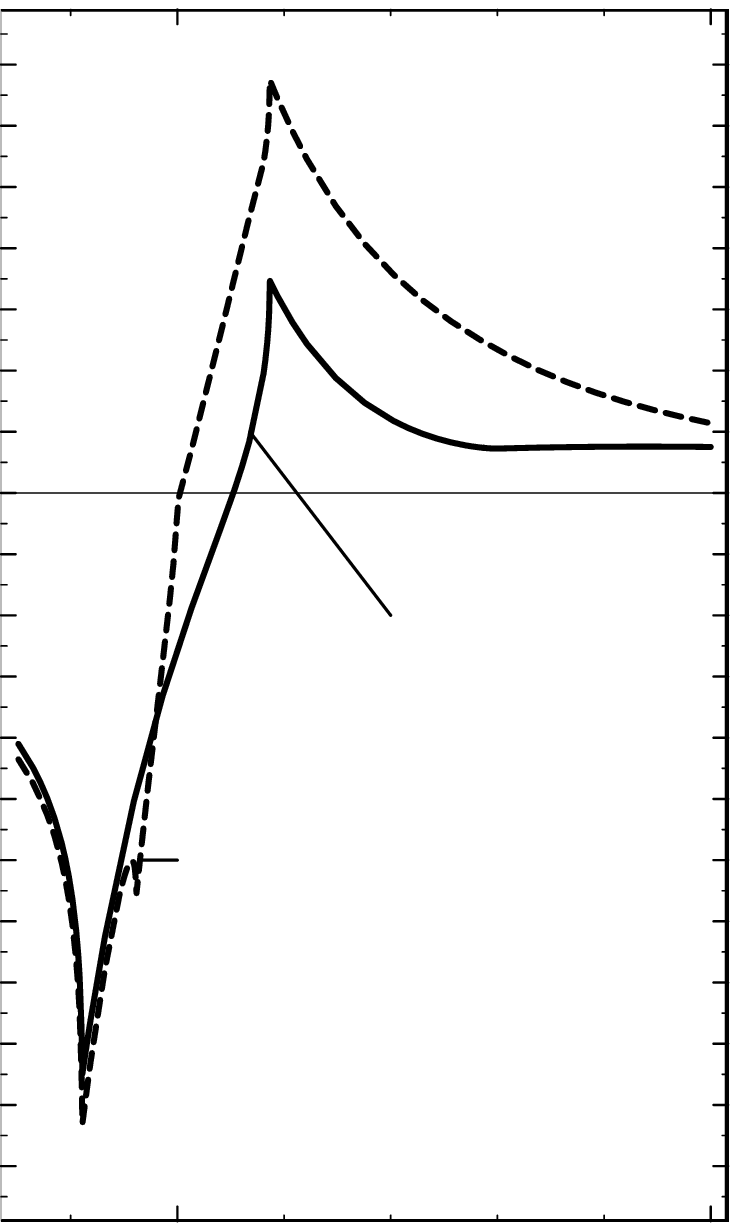,width=63mm}}}
\put(104,68.0){\makebox(0,0)[ l]{$\eRe \dgs$}}
\put( 86,46.8){\makebox(0,0)[ l]{$\eRe \dzs$}}
\put(134,110.4){\makebox(0,0)[ l]{\small  0.0006}}
\put(134, 99.8){\makebox(0,0)[ l]{\small  0.0004}}
\put(134, 89.2){\makebox(0,0)[ l]{\small  0.0002}}
\put(134, 78.6){\makebox(0,0)[ l]{\small  0}}
\put(134, 68.0){\makebox(0,0)[ l]{\small -0.0002}}
\put(134, 57.4){\makebox(0,0)[ l]{\small -0.0004}}
\put(134, 46.8){\makebox(0,0)[ l]{\small -0.0006}}
\put(134, 36.2){\makebox(0,0)[ l]{\small -0.0008}}
\put(134, 25.6){\makebox(0,0)[ l]{\small -0.0010}}
\put( 76  ,14){\makebox(0,0)[t]{\small 400}}
\put( 86.2,14){\makebox(0,0)[t]{\small 500}}
\put( 94.4,14){\makebox(0,0)[t]{\small 600}}
\put(103.6,14){\makebox(0,0)[t]{\small 700}}
\put(112.8,14){\makebox(0,0)[t]{\small 800}}
\put(122  ,14){\makebox(0,0)[t]{\small 900}}
\put(101, 2){\makebox(0,0)[b ]{\large\bf Fig.~6b}}
\end{picture}\\
\setlength{\unitlength}{1pt}
\end{center}

\clearpage

\section*{Figure Captions for $\tan\beta = 3$}

\paragraph{Figure~$\bar{2}$:} 
Dependence of the masses (in GeV) on \mbox{$\cos\varphi_{\mu}$} for 
\mbox{$M = 230$~GeV}. 
\\ \textbf{(a)} charginos: 
$m_{\tilde{\chi}^{+}_{1}}$ (full line), 
$m_{\tilde{\chi}^{+}_{2}}$ (dashed line).
\\ \textbf{(b)} neutralinos: 
$m_{\tilde{\chi}^{0}_{1}}$ (full line), 
$m_{\tilde{\chi}^{0}_{2}}$ (dashed line),
\\ \mbox{\hspace{6.3mm}}
$m_{\tilde{\chi}^{0}_{3}}$ (dotted line), 
$m_{\tilde{\chi}^{0}_{4}}$ (dashed--dotted line).

\paragraph{Figure~$\bar{3}$:} 
Dependence of the chargino contribution to $\dgs$ and $\dzs$ on 
$|\mu|$~(GeV) for the reference parameter set. 
\\ \textbf{(a)} 
\mbox{$\mIm d^{\gamma}_{\tilde{\chi}^{+}}$} for 
\mbox{$M = 230$~GeV} (full line), 
\mbox{$M = 360$~GeV} (dotted line), 
\\ \mbox{\hspace{6.3mm}}
\mbox{$\mIm d^{Z}_{\tilde{\chi}^{+}}$} for 
\mbox{$M = 230$~GeV} (dashed line), 
\mbox{$M = 360$~GeV} (dashed--dotted line). 
\\ \textbf{(b)} 
\mbox{$\eRe d^{\gamma}_{\tilde{\chi}^{+}}$} for 
\mbox{$M = 230$~GeV} (full line), 
\mbox{$M = 360$~GeV} (dotted line), 
\\ \mbox{\hspace{6.3mm}}
\mbox{$\eRe d^{Z}_{\tilde{\chi}^{+}}$} for 
\mbox{$M = 230$~GeV} (dashed line), 
\mbox{$M = 360$~GeV} (dashed--dotted line). 

\paragraph{Figure~$\bar{4}$:} 
Dependence of the neutralino contribution to $\dgs$ and $\dzs$ on 
$|\mu|$~(GeV) for the reference parameter set. 
\\ \textbf{(a)} 
\mbox{$\mIm d^{\gamma}_{\tilde{\chi}^{0}}$} for 
\mbox{$M = 230$~GeV} (full line), 
\mbox{$M = 360$~GeV} (dotted line), 
\\ \mbox{\hspace{6.3mm}}
\mbox{$\mIm d^{Z}_{\tilde{\chi}^{0}}$} for 
\mbox{$M = 230$~GeV} (dashed line), 
\mbox{$M = 360$~GeV} (dashed--dotted line). 
\\ \textbf{(b)} 
\mbox{$\eRe d^{\gamma}_{\tilde{\chi}^{0}}$} for 
\mbox{$M = 230$~GeV} (full line), 
\mbox{$M = 360$~GeV} (dotted line), 
\\ \mbox{\hspace{6.3mm}}
\mbox{$\eRe d^{Z}_{\tilde{\chi}^{0}}$} for 
\mbox{$M = 230$~GeV} (dashed line), 
\mbox{$M = 360$~GeV} (dashed--dotted line). 

\paragraph{Figure~$\bar{5}$:} 
Dependence of the chargino/neutralino contributions to 
$\dgs$ and $\dzs$ on $\varphi_{\mu}$ for the reference parameter set. 
\\ \textbf{(a)} 
\mbox{$\mIm d^{\gamma}_{\tilde{\chi}^{+}}$} for 
\mbox{$M = 230$~GeV} (full line), 
\mbox{$M = 360$~GeV} (dotted line), 
\\ \mbox{\hspace{6.3mm}}
\mbox{$\mIm d^{Z}_{\tilde{\chi}^{+}}$} for 
\mbox{$M = 230$~GeV} (dashed line), 
\mbox{$M = 360$~GeV} (dashed--dotted line). 
\\ \textbf{(b)} 
\mbox{$\eRe d^{\gamma}_{\tilde{\chi}^{+}}$} for 
\mbox{$M = 230$~GeV} (full line), 
\mbox{$M = 360$~GeV} (dotted line), 
\\ \mbox{\hspace{6.3mm}}
\mbox{$\eRe d^{Z}_{\tilde{\chi}^{+}}$} for 
\mbox{$M = 230$~GeV} (dashed line), 
\mbox{$M = 360$~GeV} (dashed--dotted line). 
\\ \textbf{(c)} 
\mbox{$\mIm d^{\gamma}_{\tilde{\chi}^{0}}$} for 
\mbox{$M = 230$~GeV} (full line), 
\mbox{$M = 360$~GeV} (dotted line), 
\\ \mbox{\hspace{6.3mm}}
\mbox{$\mIm d^{Z}_{\tilde{\chi}^{0}}$} for 
\mbox{$M = 230$~GeV} (dashed line), 
\mbox{$M = 360$~GeV} (dashed--dotted line). 
\\ \textbf{(d)}
\mbox{$\eRe d^{\gamma}_{\tilde{\chi}^{0}}$} for 
\mbox{$M = 230$~GeV} (full line), 
\mbox{$M = 360$~GeV} (dotted line), 
\\ \mbox{\hspace{6.3mm}}
\mbox{$\eRe d^{Z}_{\tilde{\chi}^{0}}$} for 
\mbox{$M = 230$~GeV} (dashed line), 
\mbox{$M = 360$~GeV} (dashed--dotted line). 

\paragraph{Figure~$\bar{6}$:} 
$\dgs$ and $\dzs$ for the reference parameter set with 
\mbox{$M = 230$~GeV}.
\\ \textbf{(a)} 
\mbox{$\mIm \dgs$} (full line), 
\mbox{$\mIm \dzs$} (dashed line)
\\ \textbf{(b)} 
\mbox{$\eRe \dgs$} (full line), 
\mbox{$\eRe \dzs$} (dashed line). 

\begin{center}
\setlength{\unitlength}{1mm}
\begin{picture}(140,120)(10,0)
\put( 7,15){\mbox{\epsfig{file=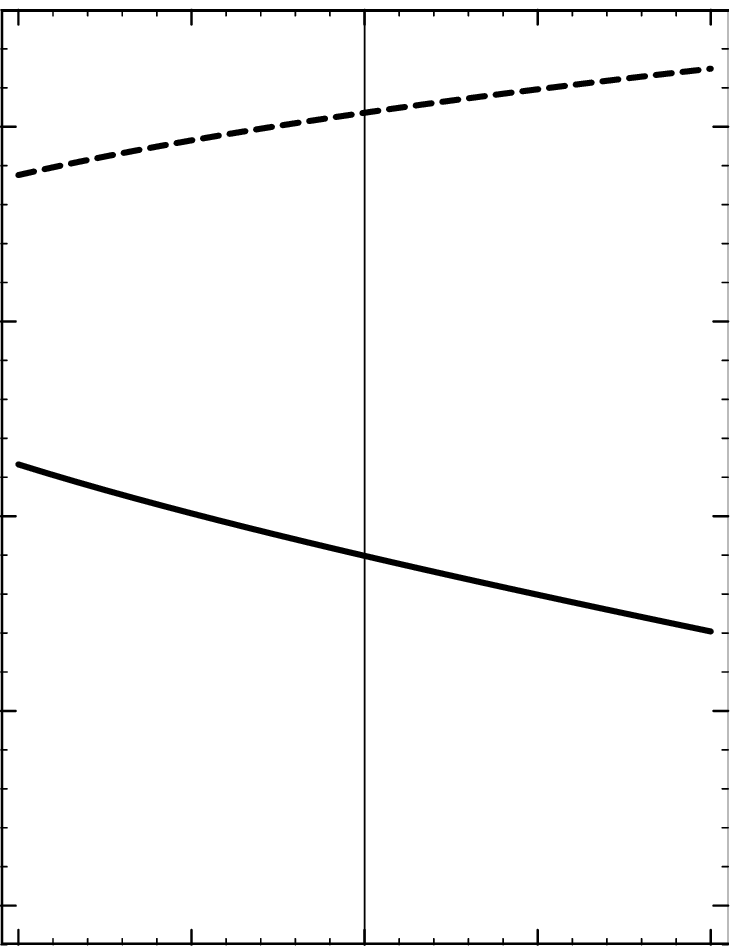,width=63mm}}}
\put( 53.5,13){\makebox(0,0)[t ]{$\cos \varphi_{\mu}$}}
\put( 35,50){\makebox(0,0)[tr]{$m_{\tilde{\chi}^{+}_{1}}$}}
\put( 35,82){\makebox(0,0)[tr]{$m_{\tilde{\chi}^{+}_{2}}$}}
\put(  6,19   ){\makebox(0,0)[ r]{\small 100}}
\put(  6,35.75){\makebox(0,0)[ r]{\small 150}}
\put(  6,52.50){\makebox(0,0)[ r]{\small 200}}
\put(  6,69.25){\makebox(0,0)[ r]{\small 250}}
\put(  6,86   ){\makebox(0,0)[ r]{\small 300}}
\put(  6,92.7 ){\makebox(0,0)[ r]{\small GeV}}
\put(  8.5,14){\makebox(0,0)[t]{\small -1}}
\put( 38.5,14){\makebox(0,0)[t]{\small  0}}
\put( 68.5,14){\makebox(0,0)[t]{\small  1}}
\put( 38, 2){\makebox(0,0)[b ]{\large\bf Fig.~$\bar{2}$a}}
%
\put(70,15){\mbox{\epsfig{file=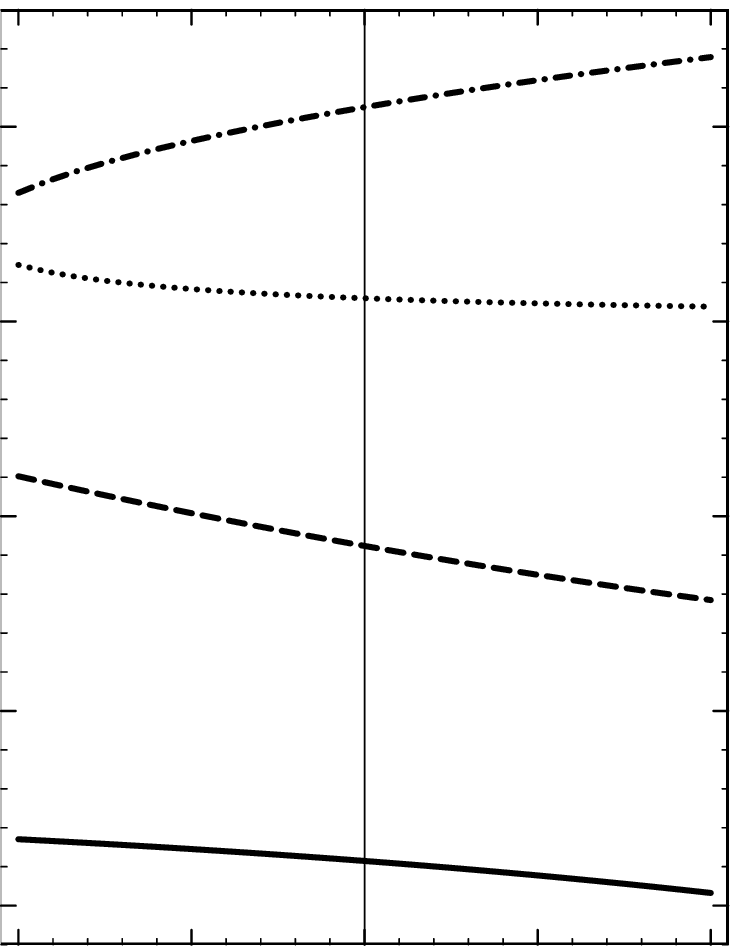,width=63mm}}}
\put(116.5,13){\makebox(0,0)[t ]{$\cos \varphi_{\mu}$}}
\put(110,26){\makebox(0,0)[tl]{$m_{\tilde{\chi}^{0}_{1}}$}}
\put(110,46){\makebox(0,0)[tl]{$m_{\tilde{\chi}^{0}_{2}}$}}
\put(110,68){\makebox(0,0)[tl]{$m_{\tilde{\chi}^{0}_{3}}$}}
\put(110,88){\makebox(0,0)[tl]{$m_{\tilde{\chi}^{0}_{4}}$}}
\put(101, 2){\makebox(0,0)[b ]{\large\bf Fig.~$\bar{2}$b}}
\put(134,19   ){\makebox(0,0)[ l]{\small 100}}
\put(134,35.75){\makebox(0,0)[ l]{\small 150}}
\put(134,52.50){\makebox(0,0)[ l]{\small 200}}
\put(134,69.25){\makebox(0,0)[ l]{\small 250}}
\put(134,86   ){\makebox(0,0)[ l]{\small 300}}
\put(134,92.7 ){\makebox(0,0)[ l]{\small GeV}}
\put( 71.5,14){\makebox(0,0)[t]{\small -1}}
\put(101.5,14){\makebox(0,0)[t]{\small  0}}
\put(131.5,14){\makebox(0,0)[t]{\small  1}}
\end{picture}\\
\setlength{\unitlength}{1pt}
\end{center}

\clearpage
\begin{center}
\setlength{\unitlength}{1mm}
%
%
\begin{picture}(140,90)(10,0)
\put( 7,15){\mbox{\epsfig{file=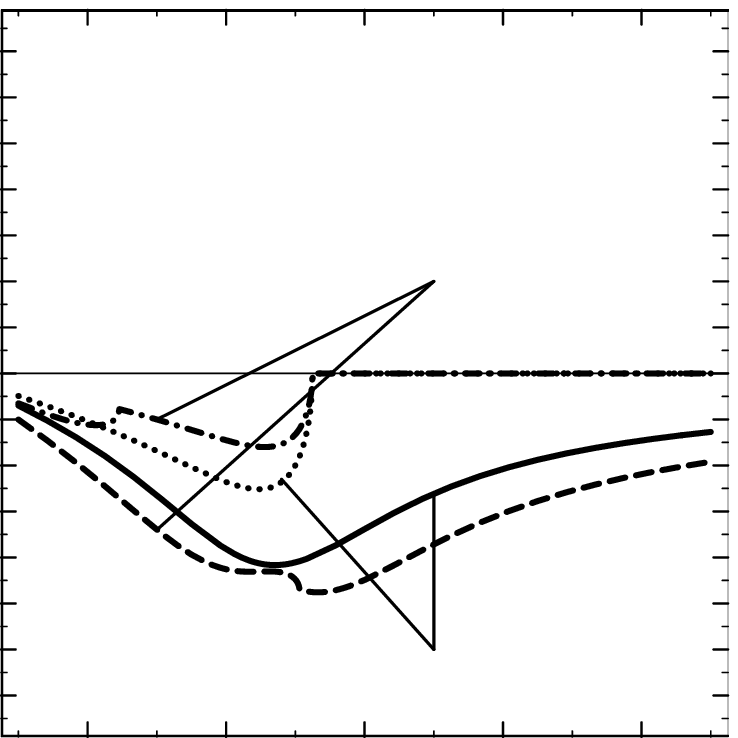,width=63mm}}}
\put( 45,54.5){\makebox(0,0)[ l]{$\mIm d^{Z}_{\tilde{\chi}^{+}}$}}
\put( 45,23.3){\makebox(0,0)[ l]{$\mIm d^{\gamma}_{\tilde{\chi}^{+}}$}}
\put(  6,70.1){\makebox(0,0)[ r]{\small  0.0006}}
\put(  6,62.3){\makebox(0,0)[ r]{\small  0.0004}}
\put(  6,54.5){\makebox(0,0)[ r]{\small  0.0002}}
\put(  6,46.7){\makebox(0,0)[ r]{\small  0}}
\put(  6,38.9){\makebox(0,0)[ r]{\small -0.0002}}
\put(  6,31.1){\makebox(0,0)[ r]{\small -0.0004}}
\put(  6,23.3){\makebox(0,0)[ r]{\small -0.0006}}
\put( 14,14){\makebox(0,0)[t]{\small 100}}
\put( 26,14){\makebox(0,0)[t]{\small 200}}
\put( 38,14){\makebox(0,0)[t]{\small 300}}
\put( 50,14){\makebox(0,0)[t]{\small 400}}
\put( 62,14){\makebox(0,0)[t]{\small 500}}
\put( 38, 2){\makebox(0,0)[b ]{\large\bf Fig.~$\bar{3}$a}}
%
\put(70,15){\mbox{\epsfig{file=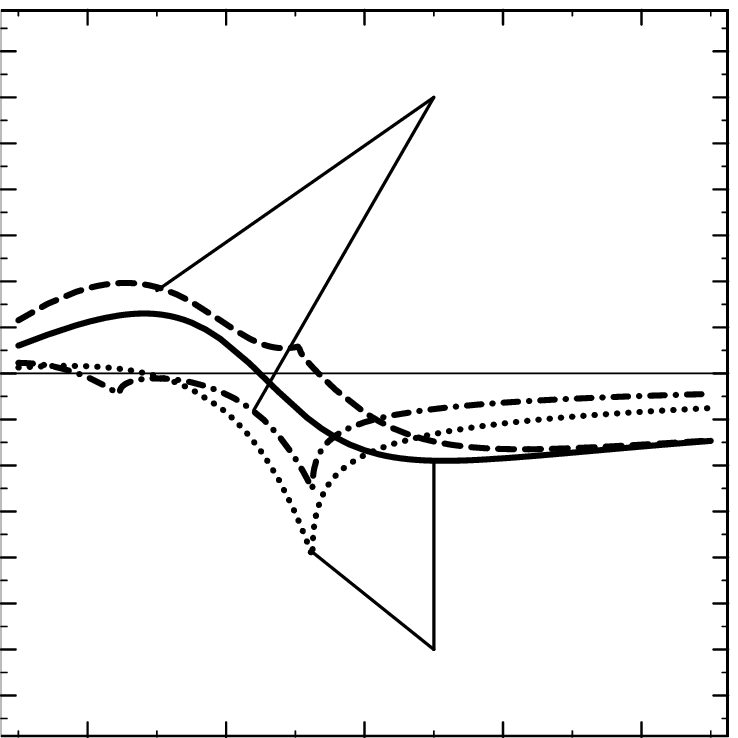,width=63mm}}}
\put( 70, 3){\makebox(0,0)[b ]{$|\mu|/$GeV}}
\put(108,70.1){\makebox(0,0)[ l]{$\eRe d^{Z}_{\tilde{\chi}^{+}}$}}
\put(108,23.3){\makebox(0,0)[ l]{$\eRe d^{\gamma}_{\tilde{\chi}^{+}}$}}
\put(134,70.1){\makebox(0,0)[ l]{\small  0.0006}}
\put(134,62.3){\makebox(0,0)[ l]{\small  0.0004}}
\put(134,54.5){\makebox(0,0)[ l]{\small  0.0002}}
\put(134,46.7){\makebox(0,0)[ l]{\small  0}}
\put(134,38.9){\makebox(0,0)[ l]{\small -0.0002}}
\put(134,31.1){\makebox(0,0)[ l]{\small -0.0004}}
\put(134,23.3){\makebox(0,0)[ l]{\small -0.0006}}
\put( 77,14){\makebox(0,0)[t]{\small 100}}
\put( 89,14){\makebox(0,0)[t]{\small 200}}
\put(101,14){\makebox(0,0)[t]{\small 300}}
\put(113,14){\makebox(0,0)[t]{\small 400}}
\put(125,14){\makebox(0,0)[t]{\small 500}}
\put(101, 2){\makebox(0,0)[b ]{\large\bf Fig.~$\bar{3}$b}}
\end{picture}\\
\setlength{\unitlength}{1pt}
\end{center}

\begin{center}
\setlength{\unitlength}{1mm}
%
%
\begin{picture}(140,90)(10,0)
\put( 7,15){\mbox{\epsfig{file=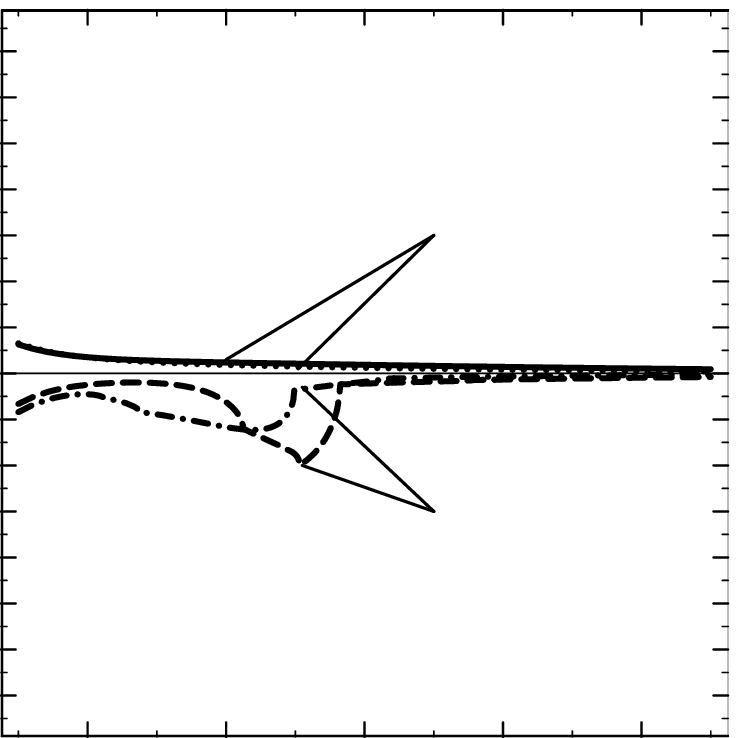,width=63mm}}}
\put( 45,33){\makebox(0,0)[ l]{$\mIm d^{Z}_{\tilde{\chi}^{0}}$}}
\put( 45,58.4){\makebox(0,0)[ l]{$\mIm d^{\gamma}_{\tilde{\chi}^{0}}$}}
\put(  6,70.1){\makebox(0,0)[ r]{\small  0.0006}}
\put(  6,62.3){\makebox(0,0)[ r]{\small  0.0004}}
\put(  6,54.5){\makebox(0,0)[ r]{\small  0.0002}}
\put(  6,46.7){\makebox(0,0)[ r]{\small  0}}
\put(  6,38.9){\makebox(0,0)[ r]{\small -0.0002}}
\put(  6,31.1){\makebox(0,0)[ r]{\small -0.0004}}
\put(  6,23.3){\makebox(0,0)[ r]{\small -0.0006}}
\put( 14,14){\makebox(0,0)[t]{\small 100}}
\put( 26,14){\makebox(0,0)[t]{\small 200}}
\put( 38,14){\makebox(0,0)[t]{\small 300}}
\put( 50,14){\makebox(0,0)[t]{\small 400}}
\put( 62,14){\makebox(0,0)[t]{\small 500}}
\put( 38, 2){\makebox(0,0)[b ]{\large\bf Fig.~$\bar{4}$a}}
\put( 70, 3){\makebox(0,0)[b ]{$|\mu|/$GeV}}
%
\put(70,15){\mbox{\epsfig{file=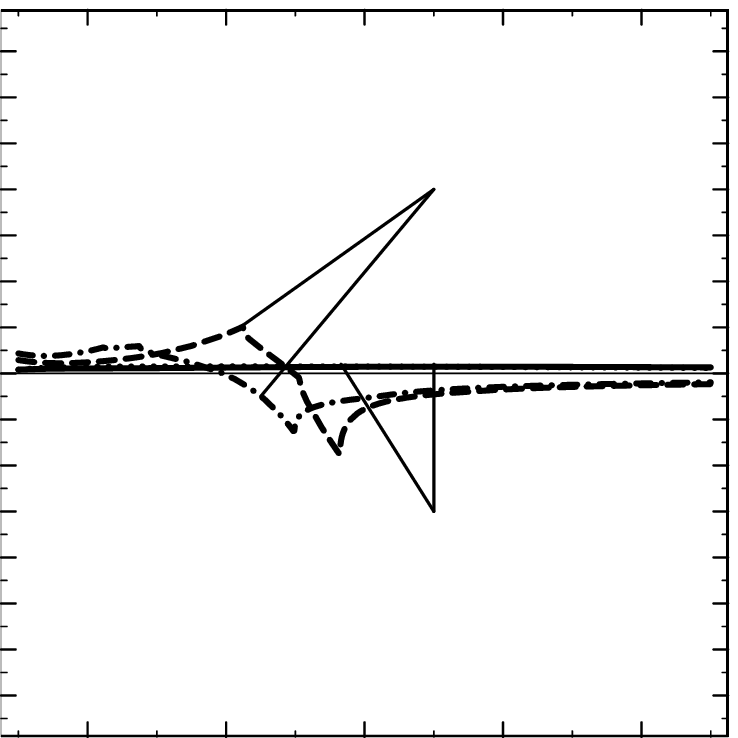,width=63mm}}}
\put(108,33){\makebox(0,0)[ l]{$\eRe d^{\gamma}_{\tilde{\chi}^{0}}$}}
\put(108,62.3){\makebox(0,0)[ l]{$\eRe d^{Z}_{\tilde{\chi}^{0}}$}}
\put(134,70.1){\makebox(0,0)[ l]{\small  0.0006}}
\put(134,62.3){\makebox(0,0)[ l]{\small  0.0004}}
\put(134,54.5){\makebox(0,0)[ l]{\small  0.0002}}
\put(134,46.7){\makebox(0,0)[ l]{\small  0}}
\put(134,38.9){\makebox(0,0)[ l]{\small -0.0002}}
\put(134,31.1){\makebox(0,0)[ l]{\small -0.0004}}
\put(134,23.3){\makebox(0,0)[ l]{\small -0.0006}}
\put( 78,14){\makebox(0,0)[t]{\small 100}}
\put( 89,14){\makebox(0,0)[t]{\small 200}}
\put(102,14){\makebox(0,0)[t]{\small 300}}
\put(113,14){\makebox(0,0)[t]{\small 400}}
\put(125,14){\makebox(0,0)[t]{\small 500}}
\put(101, 2){\makebox(0,0)[b ]{\large\bf Fig.~$\bar{4}$b}}
\end{picture}\\
\setlength{\unitlength}{1pt}
\end{center}

\begin{center}
\setlength{\unitlength}{1mm}
%
%
\begin{picture}(140,90)(10,0)
\put( 7,15){\mbox{\epsfig{file=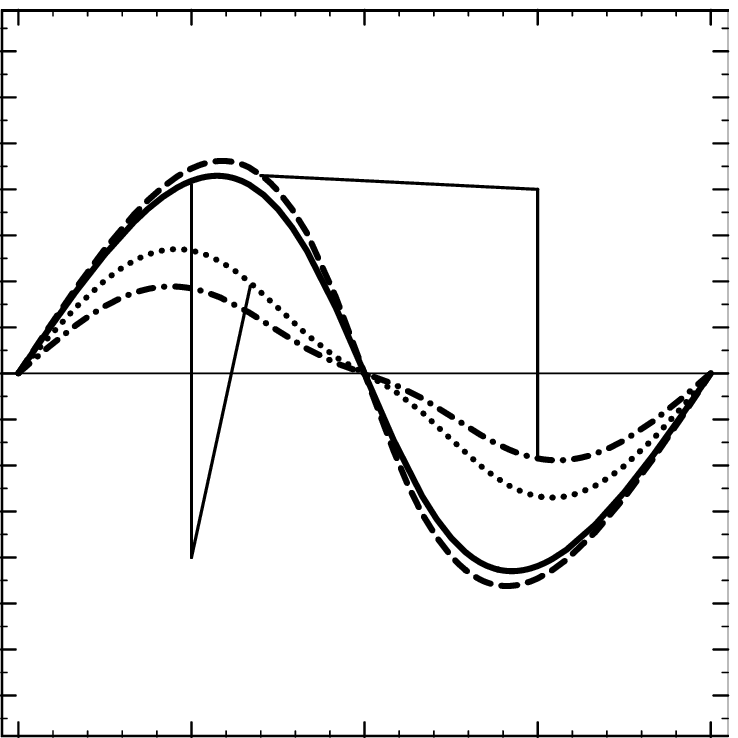,width=63mm}}}
\put( 54,62.3){\makebox(0,0)[ l]{$\mIm d^{Z}_{\tilde{\chi}^{+}}$}}
\put( 24,31){\makebox(0,0)[t ]{$\mIm d^{\gamma}_{\tilde{\chi}^{+}}$}}
\put(  6,70.1){\makebox(0,0)[ r]{\small  0.0006}}
\put(  6,62.3){\makebox(0,0)[ r]{\small  0.0004}}
\put(  6,54.5){\makebox(0,0)[ r]{\small  0.0002}}
\put(  6,46.7){\makebox(0,0)[ r]{\small  0}}
\put(  6,38.9){\makebox(0,0)[ r]{\small -0.0002}}
\put(  6,31.1){\makebox(0,0)[ r]{\small -0.0004}}
\put(  6,23.3){\makebox(0,0)[ r]{\small -0.0006}}
\put(  9  ,10){\makebox(0,0)[bl]{\small 0}}
\put( 38.5,10){\makebox(0,0)[b ]{\small $\pi$}}
\put( 68  ,10){\makebox(0,0)[br]{\small $2\pi$}}
\put( 53,  11){\makebox(0,0)[ ]{$\varphi_{\mu} \to$}}
\put( 38, 2){\makebox(0,0)[b ]{\large\bf Fig.~$\bar{5}$a}}
%
\put(70,15){\mbox{\epsfig{file=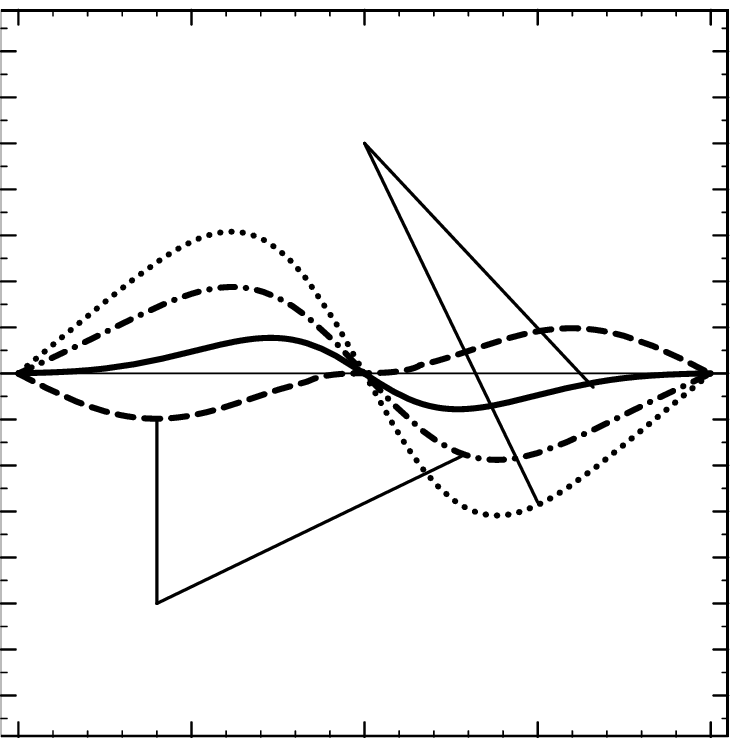,width=63mm}}}
\put( 86  ,27.1){\makebox(0,0)[t ]{$\eRe d^{Z}_{\tilde{\chi}^{+}}$}}
\put(101.5,66.5){\makebox(0,0)[b ]{$\eRe d^{\gamma}_{\tilde{\chi}^{+}}$}}
\put(134,70.1){\makebox(0,0)[ l]{\small  0.0006}}
\put(134,62.3){\makebox(0,0)[ l]{\small  0.0004}}
\put(134,54.5){\makebox(0,0)[ l]{\small  0.0002}}
\put(134,46.7){\makebox(0,0)[ l]{\small  0}}
\put(134,38.9){\makebox(0,0)[ l]{\small -0.0002}}
\put(134,31.1){\makebox(0,0)[ l]{\small -0.0004}}
\put(134,23.3){\makebox(0,0)[ l]{\small -0.0006}}
\put( 72  ,10){\makebox(0,0)[bl]{\small 0}}
\put(101.5,10){\makebox(0,0)[b ]{\small $\pi$}}
\put(131  ,10){\makebox(0,0)[br]{\small $2\pi$}}
\put(116,  11){\makebox(0,0)[ ]{$\varphi_{\mu} \to$}}
\put(101, 2){\makebox(0,0)[b ]{\large\bf Fig.~$\bar{5}$b}}
\end{picture}\\
\setlength{\unitlength}{1pt}
\end{center}

\begin{center}
\setlength{\unitlength}{1mm}
%
%
\begin{picture}(140,90)(10,0)
\put( 7,15){\mbox{\epsfig{file=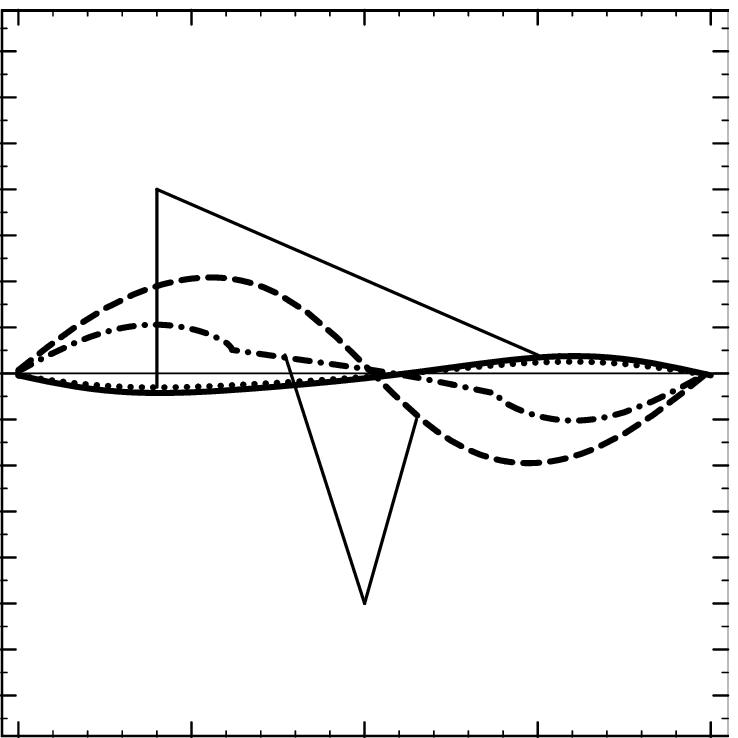,width=63mm}}}
\put( 39,26.7){\makebox(0,0)[t ]{$\mIm d^{Z}_{\tilde{\chi}^{0}}$}}
\put( 21,63.2){\makebox(0,0)[b ]{$\mIm d^{\gamma}_{\tilde{\chi}^{0}}$}}
\put(  6,70.1){\makebox(0,0)[ r]{\small  0.0006}}
\put(  6,62.3){\makebox(0,0)[ r]{\small  0.0004}}
\put(  6,54.5){\makebox(0,0)[ r]{\small  0.0002}}
\put(  6,46.7){\makebox(0,0)[ r]{\small  0}}
\put(  6,38.9){\makebox(0,0)[ r]{\small -0.0002}}
\put(  6,31.1){\makebox(0,0)[ r]{\small -0.0004}}
\put(  6,23.3){\makebox(0,0)[ r]{\small -0.0006}}
\put(  9  ,10){\makebox(0,0)[bl]{\small 0}}
\put( 38.5,10){\makebox(0,0)[b ]{\small $\pi$}}
\put( 68  ,10){\makebox(0,0)[br]{\small $2\pi$}}
\put( 53,  11){\makebox(0,0)[ ]{$\varphi_{\mu} \to$}}
\put( 38, 2){\makebox(0,0)[b ]{\large\bf Fig.~$\bar{5}$c}}
%
\put(70,15){\mbox{\epsfig{file=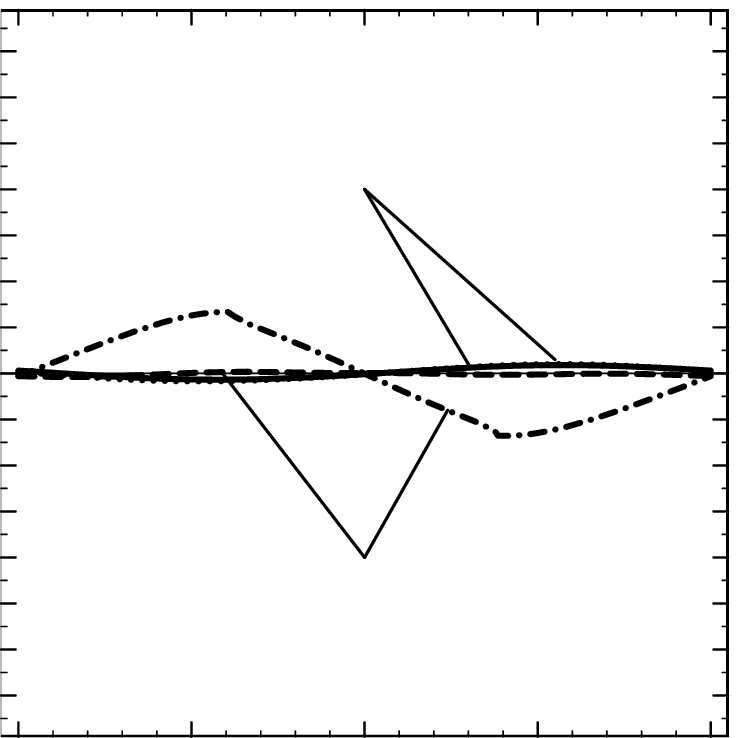,width=63mm}}}
\put(101.5,31.1){\makebox(0,0)[t ]{$\eRe d^{Z}_{\tilde{\chi}^{0}}$}}
\put(101.5,62.7){\makebox(0,0)[b ]{$\eRe d^{\gamma}_{\tilde{\chi}^{0}}$}}
\put(134,70.1){\makebox(0,0)[ l]{\small  0.0006}}
\put(134,62.3){\makebox(0,0)[ l]{\small  0.0004}}
\put(134,54.5){\makebox(0,0)[ l]{\small  0.0002}}
\put(134,46.7){\makebox(0,0)[ l]{\small  0}}
\put(134,38.9){\makebox(0,0)[ l]{\small -0.0002}}
\put(134,31.1){\makebox(0,0)[ l]{\small -0.0004}}
\put(134,23.3){\makebox(0,0)[ l]{\small -0.0006}}
\put( 72  ,10){\makebox(0,0)[bl]{\small 0}}
\put(101.5,10){\makebox(0,0)[b ]{\small $\pi$}}
\put(131  ,10){\makebox(0,0)[br]{\small $2\pi$}}
\put(116,  11){\makebox(0,0)[ ]{$\varphi_{\mu} \to$}}
\put(101, 2){\makebox(0,0)[b ]{\large\bf Fig.~$\bar{5}$d}}
\end{picture}\\
\setlength{\unitlength}{1pt}
\end{center}

\begin{center}
\setlength{\unitlength}{1mm}
%
%
\begin{picture}(140,120)(10,0)
\put( 7,15){\mbox{\epsfig{file=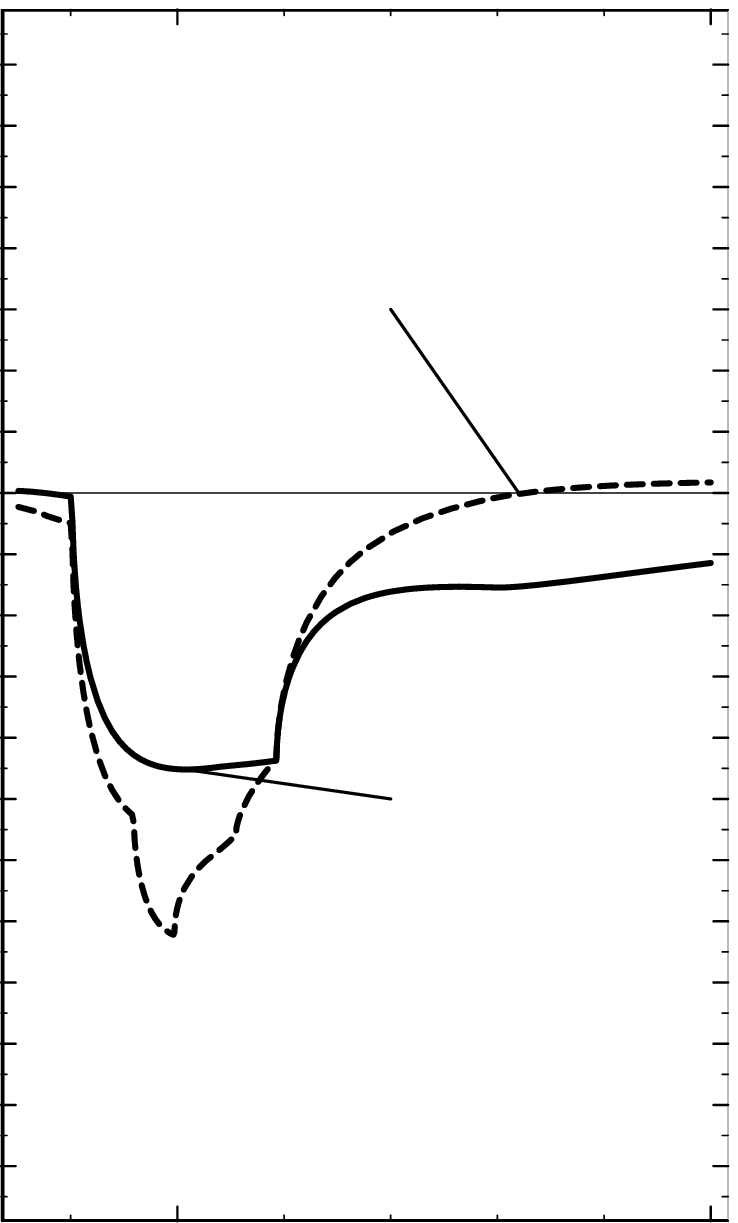,width=63mm}}}
\put( 41,94.5){\makebox(0,0)[b ]{$\mIm \dzs$}}
\put( 42,52.1){\makebox(0,0)[ l]{$\mIm \dgs$}}
\put(  6,110.4){\makebox(0,0)[ r]{\small  0.0006}}
\put(  6, 99.8){\makebox(0,0)[ r]{\small  0.0004}}
\put(  6, 89.2){\makebox(0,0)[ r]{\small  0.0002}}
\put(  6, 78.6){\makebox(0,0)[ r]{\small  0}}
\put(  6, 68.0){\makebox(0,0)[ r]{\small -0.0002}}
\put(  6, 57.4){\makebox(0,0)[ r]{\small -0.0004}}
\put(  6, 46.8){\makebox(0,0)[ r]{\small -0.0006}}
\put(  6, 36.2){\makebox(0,0)[ r]{\small -0.0008}}
\put(  6, 25.6){\makebox(0,0)[ r]{\small -0.0010}}
\put( 13  ,14){\makebox(0,0)[t]{\small 400}}
\put( 22.2,14){\makebox(0,0)[t]{\small 500}}
\put( 31.4,14){\makebox(0,0)[t]{\small 600}}
\put( 40.6,14){\makebox(0,0)[t]{\small 700}}
\put( 49.8,14){\makebox(0,0)[t]{\small 800}}
\put( 59  ,14){\makebox(0,0)[t]{\small 900}}
\put( 70, 3){\makebox(0,0)[b ]{$\sqrt{s}/$GeV}}
\put( 38, 2){\makebox(0,0)[b ]{\large\bf Fig.~$\bar{6}$a}}
%
\put(70,15){\mbox{\epsfig{file=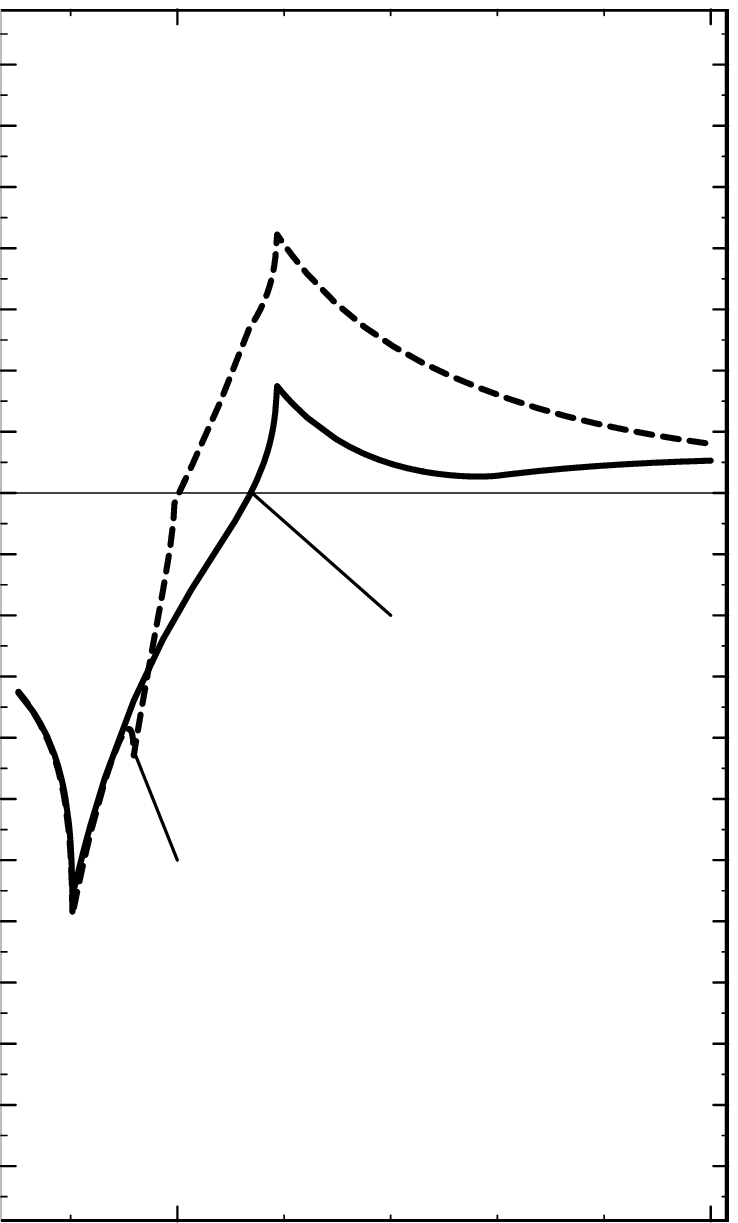,width=63mm}}}
%
\put(104,68.0){\makebox(0,0)[ l]{$\eRe \dgs$}}
\put( 86,46.8){\makebox(0,0)[ l]{$\eRe \dzs$}}
\put(134,110.4){\makebox(0,0)[ l]{\small  0.0006}}
\put(134, 99.8){\makebox(0,0)[ l]{\small  0.0004}}
\put(134, 89.2){\makebox(0,0)[ l]{\small  0.0002}}
\put(134, 78.6){\makebox(0,0)[ l]{\small  0}}
\put(134, 68.0){\makebox(0,0)[ l]{\small -0.0002}}
\put(134, 57.4){\makebox(0,0)[ l]{\small -0.0004}}
\put(134, 46.8){\makebox(0,0)[ l]{\small -0.0006}}
\put(134, 36.2){\makebox(0,0)[ l]{\small -0.0008}}
\put(134, 25.6){\makebox(0,0)[ l]{\small -0.0010}}
\put( 76  ,14){\makebox(0,0)[t]{\small 400}}
\put( 86.2,14){\makebox(0,0)[t]{\small 500}}
\put( 94.4,14){\makebox(0,0)[t]{\small 600}}
\put(103.6,14){\makebox(0,0)[t]{\small 700}}
\put(112.8,14){\makebox(0,0)[t]{\small 800}}
\put(122  ,14){\makebox(0,0)[t]{\small 900}}
\put(101, 2){\makebox(0,0)[b ]{\large\bf Fig.~$\bar{6}$b}}
\end{picture}\\
\setlength{\unitlength}{1pt}
\end{center}

%% file: paperCP.bbl
\begin{thebibliography}{99}

\bibitem{mtop}
  CDF Collaboration, F. Abe et al, 
  \emph{Phys.Rev.Lett.} \textbf{74} (1995) 2626 \\ 
  D0 Collaboration, S. Abachi et al, 
  \emph{Phys.Rev.Lett.} \textbf{74} (1995) 2632

\bibitem{CPviol} 
  W. Bernreuther and O. Nachtmann,  
  \emph{Phys.Lett.} \textbf{B268} (1991) 424 \\  
  A. Brandenburg, J. P. Ma, R. M\"unch, O. Nachtmann,   
  \emph{Z. Phys.} \textbf{C51} (1991) 225 \\ 
  D. Atwood and A. Soni, 
  \emph{Phys.Rev.} \textbf{D45} (1992) 2405,
  \textbf{hep-ph} / 9609418 \\ 
  W.Bernreuther, O.Nachtmann, P.Overmann, T.Schr\"oder,  
  \emph{Nucl.Phys.} \textbf{388} (1992) 53\\ 
  C. R. Schmidt  
  \emph{Phys.Lett.} \textbf{B293} (1992) 111 \\  
  C. R. Schmidt and M. E. Peskin,  
  \emph{Phys.Rev.Lett.} \textbf{69} (1992) 410 \\  
  W. Bernreuther and A. Brandenburg,  
  \emph{Phys.Lett.} \textbf{B314} (1993) 104; \\  
  D. Chang, Wai-Yee Keung and I. Phillips,    
  \emph{Nucl.Phys.} \textbf{B408} (1993) 286 \\  
  B. Grzadkowski and W. Keung,  
  \emph{Phys.Lett.} \textbf{B319} (1993) 526 \\ 
  F. Cuypers and S. Rindani, 
  \emph{Phys.Lett.} \textbf{B343} (1995) 333 \\ 
  P. Poulose and S. Rindani, 
  \emph{Phys.Lett.} \textbf{B349} (1995) 379 

\bibitem{dipgl} 
  E. Christova and M. Fabbrichesi,  
  \emph{Phys.Lett.} \textbf{B315} (1993) 338\\ 
  B. Grzadkowski,  
  \emph{Phys.Lett.} \textbf{B305} (1993) 384 \\ 
  W. Bernreuter, P. Overmann,  
  \emph{Z.Phys.} \textbf{C61} (1994) 599
  
\bibitem{we} 
  A. Bartl, E. Christova, W. Majerotto, 
  \emph{Nucl.Phys.} \textbf{B460} (1996) 235; erratum 
  \emph{Nucl.Phys.} \textbf{B465} (1996) 365;

\bibitem{C.J.} 
  C.J.-C. Im, G.L. Kane and P.J. Malde, 
  \emph{Phys.Lett.} \textbf{B317} (1993) 454

\bibitem{Kane}
  H.E. Haber, G.L. Kane,
  \emph{Phys.Rep.} \textbf{177} (1985) 75 \\
  J. Gunion and H. E. Haber, 
  \emph{Nucl.Phys.} \textbf{B272} (1986) 1

\bibitem{Dugan}  
  M. Dugan, B. Grinstein, L. Hall, 
  \emph{Nucl.Phys.} \textbf{B255} (1985) 413 \\ 
  W. Bernreuter, M. Suzuki 
  \emph{Rev.Mod.Phys.} \textbf{63} (1991) 313 \\
  W. Hollik, J.I. Illana, S. Rigolin, D. St\"ockinger, 
  \textbf{hep-ph} / 9711322

\bibitem{Osh1} 
  Y. Kizukuri and N. Oshimo, 
  \emph{Phys.Rev.} \textbf{D45} (1992) 1806

\bibitem{SingW}
  J.M. Ortega,
  {\em Matrix Theory} (Plenum Press, New York 1987)

\bibitem{Garisto}
  R. Garisto, J.D. Wells, 
  \emph{Phys.Rev.} \textbf{D55} (1997) 1611

\bibitem{new}
  C. Hamzaoui, M. Pospelov, R. Roiban,
  \emph{Phys.Rev.} \textbf{D56} (1997) 4295

\bibitem{Osh2} 
  Y. Kizukuri and N. Oshimo,
  \emph{Phys.Rev.} \textbf{D46} (1992) 3025

\bibitem{sabine}
  S. Kraml, H. Eberl, A. Bartl, W. Majerotto, W. Porod,
  \emph{Phys.Lett.} \textbf{B386} (1996) 175 

\bibitem{BilPet}
  S. M. Bilenky, S. T. Petcov, 
  \emph{Rev.Mod.Phys.} \textbf{59}, No. 3, Part 1 (1987) 671

\bibitem{PaVe}
  G. Passarino and M. Veltman, 
  \emph{Nucl.Phys.} \textbf{B160} (1979) 151

\bibitem{Denner}
  A. Denner,
  \emph{Progress of Physics} \textbf{41} (1993) 4, 307

\end{thebibliography}
